\documentclass[10pt,twocolumn]{article}

\usepackage[
  paper=a4paper,
  left=20mm,
  right=20mm,
  top=25mm,
  bottom=25mm
]{geometry}

\usepackage{times}            
\usepackage{mathptmx}         
\usepackage{amsmath,amssymb}

\usepackage{graphicx}
\usepackage{xcolor}
\usepackage{url}

\usepackage{titlesec}
\titlespacing*{\section}{0pt}{*1.8}{*0.8}
\titlespacing*{\subsection}{0pt}{*1.4}{*0.6}

\usepackage[font=small]{caption}

\usepackage{graphicx}
\usepackage{amsmath,amssymb}
\usepackage{url}
\usepackage{xcolor}
\usepackage[T1]{fontenc}
\usepackage{dfadobe}  

\usepackage{cite}  
\usepackage{subcaption}
\usepackage{xspace} 
\newcommand{\etal}{et al.\@\xspace}
\usepackage{placeins}

\title{The Truth, the Whole Truth, and Nothing but the Truth:\\
Automatic Visualization Evaluation from Reconstruction Quality}

\author{
Roxana Bujack \and
Li-Ta Lo \and
Ethan Stam \and
Ayan Biswas \and
David Rogers\\[4pt]
}

\date{Los Alamos National Laboratory, USA} 

\begin{document}
\maketitle

\maketitle
\begin{abstract}
 Recent advances in AI enable the automatic generation of visualizations directly from textual prompts using agentic workflows. However, visualizations produced via one-shot generative methods often suffer from insufficient quality, typically requiring a human in the loop to refine the outputs. Human evaluation, though effective, is costly and impractical at scale. To alleviate this problem, we propose an automated metric that evaluates visualization quality without relying on extensive human-labeled datasets. Instead, our approach uses the original underlying data as implicit ground truth. Specifically, we introduce a method that measures visualization quality by assessing the reconstruction accuracy of the original data from the visualization itself. This reconstruction-based metric provides an autonomous and scalable proxy for thorough human evaluation, facilitating more efficient and reliable AI-driven visualization workflows.


\end{abstract}  
\section{Introduction}

Recent advances in artificial intelligence have made it possible to automatically generate visualizations directly from textual prompts~\cite{maddigan:2023:chat2vis,mallick:2025:chatvis}. Despite these promising capabilities, visualizations created using one-shot generative AI methods are frequently of inadequate quality. Effective visualization must provide users with a truthful mental representation or \emph{mental map} of the underlying dataset to facilitate exploration, insight, discovery, and knowledge formation~\cite{ware2019information, tufte1983visual, shneiderman2003eyes, kindlmann2014algebraic}. A visualization that fails to convey data faithfully can mislead users or, in extreme cases, constitute ``lying'' with visualization~\cite{tufte1983visual, rogowitz1996not, kindlmann2014algebraic, chen2014visualization}.
To overcome these limitations, human-in-the-loop processes are commonly employed to iteratively select and refine appropriate visualizations. However, such human evaluation is expensive, time-consuming, and impractical at large scale. 

The generation of everyday images shows how high-quality results depend on huge training datasets~\cite{sun2017revisiting, dosovitskiy2020image}, whose size exceeds what exists for scientific visualizations, such as papers, images, user studies, and experiments, by many orders of magnitude.
Consequently, there is a clear need for automated quality metrics capable of assessing visualization quality without extensive human-labeled data.

Many existing automated metrics rely on heuristics derived from perceptual psychology or information theory, such as salience-based metrics~\cite{janicke2010salience}, viewpoint entropy~\cite{bordoloi2005view,vazquez2001viewpoint}, or structural similarity indices~\cite{wang2005structural}. Though effective for specific datasets, their suitability for general exploratory analysis remains limited. 

In this paper, we address visualization quality assessment specifically within the context of exploratory data analysis. We propose a novel autonomous metric that evaluates the fidelity of visualizations by leveraging the available inherent ground truth---the original data itself. Inspired by Jänicke et al.'s concept of visual reconstructability in 2D vector fields~\cite{janicke2011visual}, we introduce a workflow that samples the visualization parameter space and selects the visualization from which the original data can be reconstructed with the greatest accuracy. Specifically, we are the first to extend this concept to 3D. We employ Neural Radiance Fields (NeRF)~\cite{mildenhall2021nerf} to reconstruct 3D data from 2D visualizations.


We argue for using reconstruction accuracy as a metric that allows automatic evaluation of visualization quality for the task of data exploration. We showcase examples of isocontours and colormapping, which are two of the most common visual representations for scalar data. Even though we limit our evaluation to the scientific visualization of scalar fields, we make an argument for the general case of visualization of arbitrary data and arbitrary visualization metaphors. If the visualization satisfies Kindlmann's Principle of Visual-Data Correspondence~\cite{kindlmann2014algebraic}, then it tells us the whole truth (because changes in the data result in changes in the visualization) and nothing but the truth (because changes in the visualization correspond to changes in the data). We argue that choosing visualizations that maximize the truthful reconstruction of the original data is a promising direction for satisfying this principle, and that can be automatically evaluated in AI workflows.

We follow a very simple approach to data reconstruction to illustrate that our approach works in principle. We think of it mostly as how a computer would reconstruct the data based on the visualization. This is not intended to be the final method of reconstruction, nor the final metric of visualization quality. The ultimate goal is to build a model of human perception and cognition that is trained on experiments on humans and mimics their process of building a mental image of the underlying data. The first step of this ambitious goal is to be able to reconstruct the data from the visualization and this is what we focus on in this paper. We believe that our approach can in the future be extended to incorporate specific properties, such as limits and biases of human perception, to perform tasks, to interpret exaggerated or abstracted representations, and to apply domain knowledge. 

\section{Disclaimer}\label{s:disclaimer}
This work intends to suggest a first approximation of a paradigm for applying large-scale deep learning to visualization despite its inherent limited access to labeled data. It is very far from being a fully functional technique that can replace a human in the loop. We seek to be as forthcoming as possible about its current shortcomings and hope that by doing so, we enable the community to develop it into a fully functional tool in the future.

\textbf{Reconstructability is not Sufficient.}
While our reconstruction-based metric provides an autonomous, scalable approach to evaluating visualization quality, it should not be understood to be a complete measure of visualization effectiveness. Visualization quality extends far beyond the accurate recovery of underlying data. High-quality visualizations often highlight or emphasize the foreground, important patterns, or specific structures. Our approach could be complemented by mechanisms that evaluate importance~\cite{matzen2017data}.

\textbf{Reconstructability is not necessary.}
Visualization quality is much more than a faithful representation of the data; sometimes it is even in conflict with it. Abstraction, exaggeration, storytelling, cognitive efficiency, and clarity can matter more than raw data fidelity. Some of the best visualizations intentionally distort, filter, or simplify data to reveal patterns or reduce complexity~\cite{tufte1983visual}. We argue that a refined model of human perception, for which our approach is only a first step, will be able to interpret such types of visualizations, too, to build a mental model of the underlying data, just as humans are able to interpret exaggerations and abstractions.
                
\textbf{Human factors are critical.}
Moreover, this method assumes that ``truthful'' visualizations are those from which the data can be reconstructed with minimal loss, which neglects the broader human and contextual factors that govern effective visual communication. True evaluation of visualization quality must integrate human judgment and domain knowledge. A complete AI model of human perception is currently not yet available. Hence, this work represents a first step toward unsupervised, reinforcement-learning-based visualization evaluation, providing at least some currently-available proxy in the urgent need for scalable, automated feedback loops. 

\textbf{The tasks are critical.}
Just because we assume that no specific task is at hand in our scenario, we do not claim that this case is a very common one for visualization designers. We believe, though, that for many tasks the observers first build a mental representation of the data and then perform the tasks with its help. In that sense, the reconstruction is the basic step that we tackle first, on top of which we would need to build task-specific metrics.

\textbf{Specific heuristics excel in specific cases.}
We do not claim that our metric is better than existing heuristic evaluations, but these heuristics only work for specific data types or specific visualization metaphors. Because they are specifically designed for them, we expect them to excel. In contrast, our metric can be generalized to any data and visualization type, and on top of that, it can take into account cross-relations between several variables influencing the quality of a visualization, such as interactions between color, isovalue, and viewing angles, all at the same time. In contrast, the global optimum may not be found applying each heuristic separately. Also, some visualization variables do not have heuristics, e.g., there is no agreed-upon heuristic for how transparent a visual element should be.

\subsection{Summary}
We make the claim that a well-designed and well-trained model of human perception and cognition would be able to reconstruct the underlying data from a visualization even if it involves abstractions, simplifications, and exaggerations, but not in a one-to-one manner. But if the human visual bandwidth is properly captured by the model, we can expect a simpler visualization to allow capturing the overall structure of the data better than an overwhelming amount of clutter and detail that can only be captured by a computer with sufficient memory and compute power.

The one place where the construction of a mental model of the data seriously deviates from visualization quality, even if all human factors are properly accounted for, are task-specific visualizations. We will treat those specifically in future work but consider the mental model as a necessary foundation for learning the successful completion of tasks.

\section{Related Work}

\textbf{Visualization metrics} quantify effectiveness via perceptual clarity, cognitive efficiency, and analytical utility. 
Bertini et al.\ give an overview of metrics for high-dimensional data (clustering, correlation, outliers)~\cite{bertini2011quality}. 
Behrisch et al.\ systematize metrics addressing clutter and saliency~\cite{behrisch2018quality}. 
Wang et al.\ propose structural metrics (e.g., SSIM)~\cite{wang2005structural}. 
Bolte and Bruckner emphasize entropy and mutual information~\cite{bolte2020measures}. 
Kindlmann and Scheidegger use algebraic invariants for design~\cite{kindlmann2014algebraic}. 
Wong et al.\ compare volume-rendered images~\cite{wong2006perceptual}. 
Chen and Golan optimize via information theory~\cite{chen2015may}. 
The question of how to measure the quality of a visualization is one of the core challenges of visualization research and has interested researchers for a long time. 
 
Most proposed metrics are specific to a type of data, rule-based, and empirical. 
We suggest a metaphor that is data-agnostic.

\textbf{Data reconstruction as a quality metric} measures data preservation. 
Mackinlay et al.\ examine the expressiveness and effectiveness of a graphical layout in relation to how much information is recoverable~\cite{mackinlay1988applying}. Jänicke et al.\ define visual reconstructability~\cite{janicke2011visual}. Bramon et al.\ treat visualization as an information channel~\cite{bramon2013information}. Decatur and Krishnan automate data extraction~\cite{decatur2021vizextract}. Ye et al.\ leverage neural embeddings~\cite{ye2022visatlas}. Chen et al.\ highlight accurate reconstruction~\cite{chen2014visualization}.

How much and how accurately information about the underlying data can be recovered from a visualization is a recurring theme in visualization quality research. To date, no 3D reconstruction has been attempted.

\textbf{Viewpoint selection} finds optimal camera angles. Vázquez et al.\ introduce viewpoint entropy~\cite{vazquez2001viewpoint}. Bordoloi and Shen apply entropy for volume rendering~\cite{bordoloi2005view}. Tao et al.\ unify streamline and viewpoint selection~\cite{tao2012unified}. Takahashi et al.\ emphasize salient features~\cite{takahashi2005feature}. Marsaglia et al.\ align entropy with user preferences~\cite{marsaglia2021entropy}.

All of these approaches treat viewpoint selection as separate from the visualization design. Cross-relations between the two are not taken into account by any of them, in contrast to our approach. 

\textbf{Colormap quality metrics} are heuristics that guide the design of good colormaps~\cite{bujack2017good}. The most popular ones are order~\cite{wainer1980empirical, bujack2018ordering}, discriminative power~\cite{tajima1983uniform, mittelstadt2014revisiting}, uniformity~\cite{meyer1980perceptual, levkowitz1992design}, and smoothness~\cite{nardini2021automatic}. A colormap should also be robust with respect to vision deficiencies~\cite{rheingans2000task}, contrast effects~\cite{ware1988color}, and shading~\cite{moreland2009diverging}.

Many of these design rules are conflicting, and the question of a good trade-off has so far been the responsibility of the user. An objective metric allows for an automatic choice tailored to the data and the visualization metaphor.

\textbf{Meaningful isovalue selection} significantly affects the visualization of scalar fields. Kindlmann and Durkin proposed semi-automatic generation of transfer functions, identifying significant isovalues based on gradient magnitudes and second derivatives~\cite{kindlmann1998semi}. Carr et al.\ used contour trees to identify significant topological features~\cite{carr2004simplifying,carr2010flexible}. Bruckner and Möller developed isosurface similarity maps, employing mutual information between geometric properties~\cite{bruckner2010isosurface}.

These approaches are mostly heuristic and do not consider cross-relations with viewing conditions, such as angle, shading, or color.

\textbf{Neural radiance fields (NeRFs)}
were proposed by Mildenhall et al.\ which use a neural network to represent a function that maps $x, y, z$ coordinates and view direction $\theta, \phi$ to RGB colors and density $\sigma$~\cite{mildenhall2021nerf}. The radiance field is then volume-rendered from camera pose and compared with the ground-truth image in the training set. The difference is then back-propagated to train the neural network. Once trained, images from novel view angles can be synthesized. The original NeRF required long training times, which was greatly improved by Instant-NGP~\cite{mueller2022instant}. More recently, a diffusion model was used to generate a consistent set of images and camera poses by using only a single image as input, which are then fed to a NeRF model to create a 3D scene from a single image~\cite{gao2024cat3d}. 

NeRFs as originally proposed also required per-scene optimization which makes them difficult to be generalized. Recent research large reconstruction models (LRM)~\cite{hong2023lrm, wei2024meshlrm, xiang2025structured} proposed to pre-train transformers ~\cite{vaswani2017attention} with large scale 3D datasets to produce NeRF models. Once trained those LRM can quickly generate a trained NeRF model based on unseened input images.

To date, these techniques have not been applied to visualization quality evaluation.

\textbf{AI-generated visualization} is a rapidly expanding ecosystem of AI-powered visualization assistants. 
It ranges from language-to-visualization systems such as Chat2VIS~\cite{maddigan:2023:chat2vis}, ChatVis~\cite{mallick:2025:chatvis, peterka2025chatvis}, and ncNet~\cite{luo2021_ncNet}, to mixed-initiative analytic agents such as LightVA~\cite{zhao:2024:lightva} and PhenoFlow~\cite{kim:2024:PhenoFlow}, and further to multi-agent chart-refinement pipelines such as MatPlotAgent~\cite{yang:2024:matplotagent}, VisPath~\cite{vispath}, and PlotGen~\cite{plotgen}. 
In the context of 3D scientific visualization, VizGenie introduces a self-improving, agentic workflow that couples pre-defined VTK-based tools with LLM-generated modules and domain-tuned vision models to support feature-centric, natural-language exploration of HPC-scale volumetric datasets~\cite{vizgenie}. Complementing this, Natural Language Interaction for Volume Visualization (NLI4VolVis) integrates editable 3D Gaussian splatting, semantic segmentation, and a multi-agent LLM controller to enable open-vocabulary querying, editing, and stylization of volumetric scenes via free-form dialogue~\cite{nli4volvis}.
 
Collectively, these tools suggest considerable promise for LLM-driven automation of visualization workflows. 
So far, these methods are generative. 
They do not include automatic quality control but rely on the user to accept or reject a visualization.

\section{Workflow}
Large language models (LLMs) have reached a level of sophistication that they can generate complex visualization scripts, such as Python code or ParaView state files, directly from natural language prompts. These scripts, when executed, produce visualizations that often align remarkably well with user intent. 
However, these outputs are rarely perfect. They typically require careful user evaluation and iterative refinement, a process that is both time-consuming and subjective. To accelerate and systematize this workflow, we propose an agentic visualization pipeline where the first round of evaluation is delegated to the system itself rather than the user.
Visualization is typically a human-in-the-loop task. Places in which no human is available, such as in-situ visualization and AI generated visualizations, are tricky. Our goal is not to remove the human, but to reduce cognitive load by prioritizing candidate outputs for inspection. This is achieved through a structured and partially autonomous loop:
\begin{itemize}\setlength{\itemsep}{0pt} 
\item Translate user prompts into a visualization script; 
\item Automatically identify tunable parameters in the script;
\item Systematically generate candidate visualizations by sweeping the parameter space (and camera viewpoints in 3D); and,
\item Evaluate candidates using proxy quality metrics, then select the most promising one based on minimal reconstruction error. 
\end{itemize}

Reconstruction error is a flexible, content-aware proxy for visualization quality. It allows us to evaluate output fidelity without relying on human-provided labels. This is particularly useful when combining different visualization metaphors, such as colormapping, isosurfacing, or vector field glyphs, each of which traditionally relies on its own task-specific quality metric, e.g., perceptual uniformity, salient feature coverage, or viewpoint entropy. Reconstruction quality provides a unified, data-driven optimization framework that works for all of their combinations.
At present, tailored algorithms for different visualization techniques are employed, which limits the range of visualization metaphors to the set that we have anticipated, i.e., colormapping or isocontouring, but our long-term goal is to integrate them into a single adaptive agent trained across modalities and tasks. We see two ways forward to achieve this goal. First, we let the agent write the decoder together with the encoder. As it chooses a visualization metaphor, it also specifies how to interpret it in the reconstruction. Second, we train a general model of human perception, which will infer the visualization metaphor from the image and apply its own interpretation to what the underlying data looks like. The advantage of the latter is that it includes an automatic visualization literacy check that would avoid exotic custom metaphors that require a long learning phase. Its downside is the complexity and cost of training such a model.

\begin{figure}
    \centering
    \includegraphics[width=\linewidth]{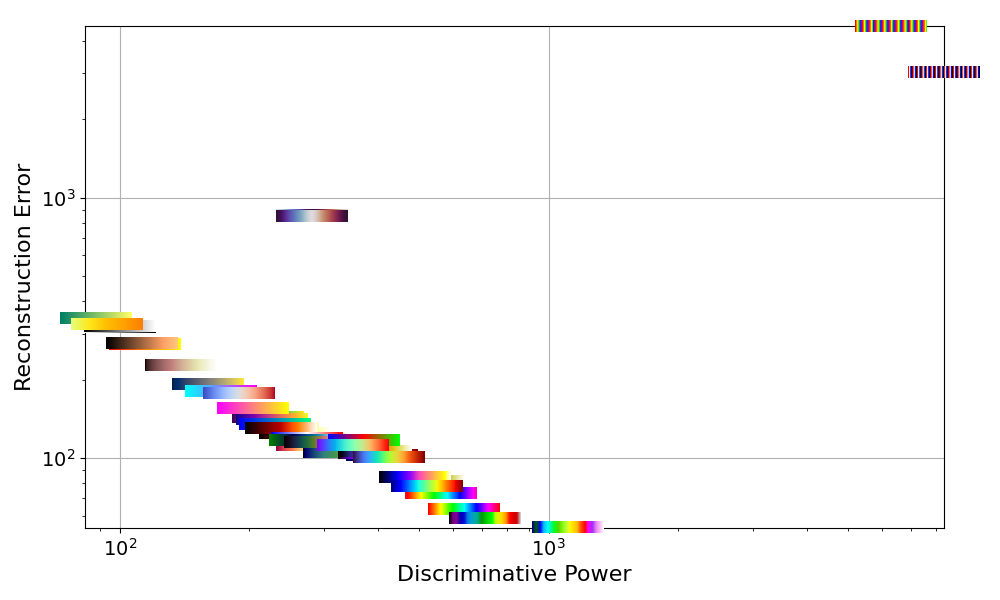}
    \caption{Log-log plot showing the strong correlation between discriminative power and reconstruction error for colormaps that preserve legend-based order.}
    \label{fig:legend}
\end{figure}

\begin{figure}[ht]
    \centering
    \subcaptionbox{\texttt{gist\_ncar}, the best performer, has the highest discriminative power among all legend-based ordered maps.}{\includegraphics[width=\linewidth]{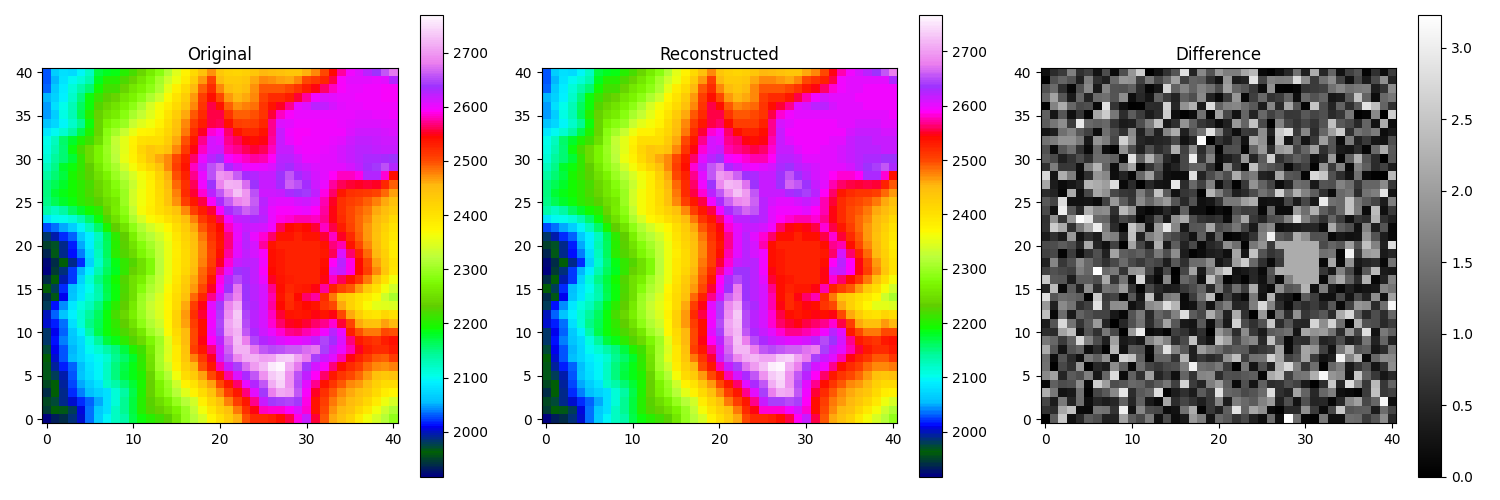}}\hfill
    \subcaptionbox{The perceptually uniform \texttt{gray} colormap exhibits low discriminative power. Its poor performance stems from the difficulty observers have in distinguishing similar shades.}{\includegraphics[width=\linewidth]{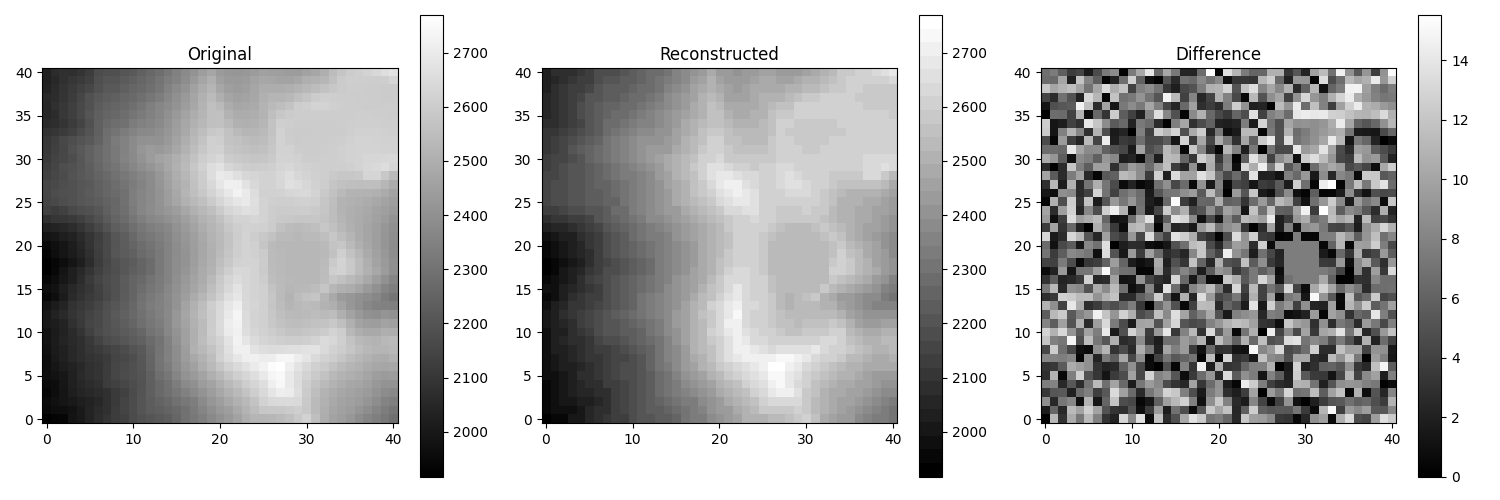}}\hfill
    \subcaptionbox{The cyclic \texttt{twilight} colormap does not preserve order. White appears both at the minimum and maximum, leading to ambiguous interpretations.}{\includegraphics[width=\linewidth]{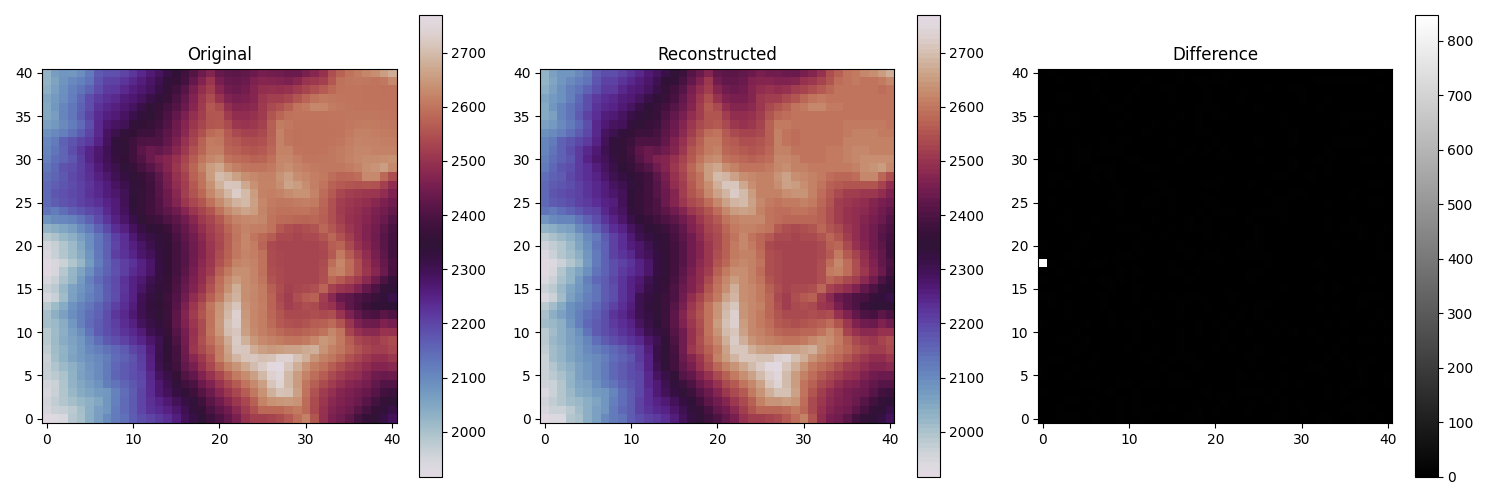}}\hfill
    \subcaptionbox{\texttt{prism}, the worst performer, violates order throughout the colormap, making it difficult to associate colors with scalar values.}{\includegraphics[width=\linewidth]{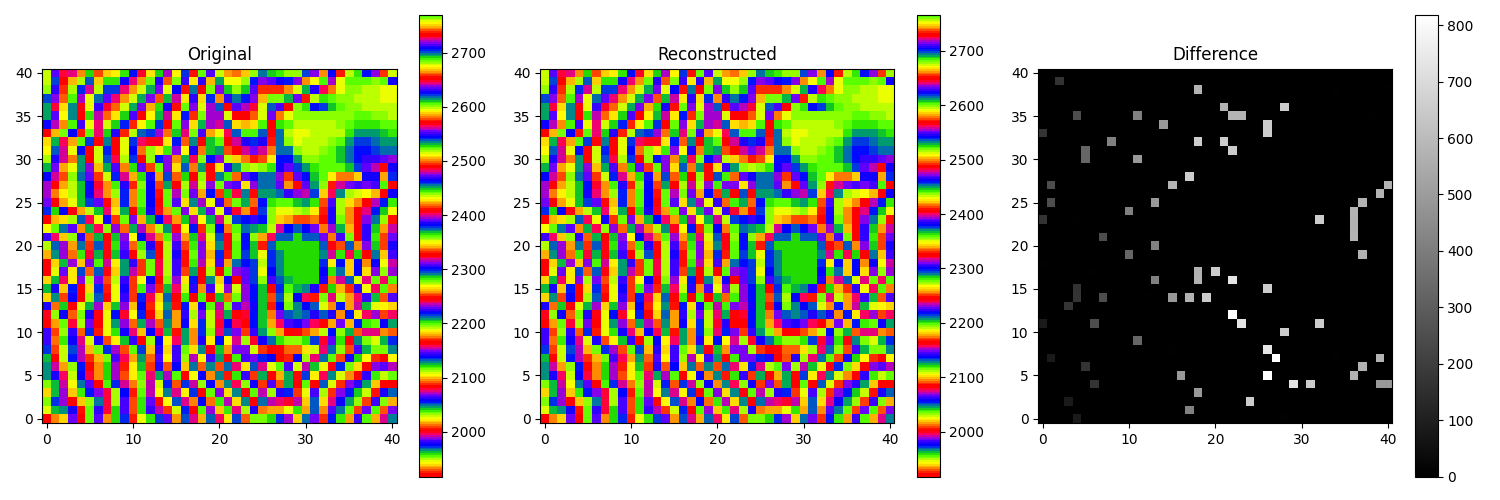}}\hfill
    \caption{Examples of reconstruction and error patterns for various colormaps. The visualizations highlight the importance of both order preservation and discriminative power for supporting accurate mental models.}
    \label{fig:legendExamples}
\end{figure}


\section{2D Colormapping}
Colormapping is one of the most widely used techniques for visualizing two-dimensional scalar fields. These visualizations are typically accompanied by a legend containing a colorbar and labeled tick-marks indicating reference data values.

\subsection{The Mental Image}
We assume that when interpreting a colormap, observers identify a pixel's color and match it to the closest color in the legend. By interpolating between labeled ticks, they estimate the corresponding scalar value. Human perceptual studies suggest that one just-noticeable difference corresponds to approximately $2.9\Delta E_{2000}$~\cite{luo2001development, gaurav2017digital}.

\subsection{Reconstruction}
To mimic this perceptual decoding process, we compute the total perceptual arc length of each colormap and divide it into uniform intervals of $2.9\Delta E_{2000}$. During reconstruction, each pixel color is compared to these binned intervals using the $\Delta E_{2000}$ color distance metric. The reconstructed scalar value is assigned as the center value of the closest matching bin.

\subsection{Evaluation}
We evaluated all Matplotlib colormaps using a subset of PyVista's elevation dataset, restricted to the region $[600, 800] \times [500, 700]$ and subsampled at a $5 \times 5$ resolution.

As expected, the worst-performing colormaps were those that violate legend-based order~\cite{bujack2018ordering}, such as \texttt{prism}, \texttt{flag}, and the cyclic colormaps. These colormaps confuse the viewer by mapping the same color to multiple values or introducing large perceptual jumps that don't correspond to scalar changes.
Conversely, the best-performing colormap, \texttt{gist\_ncar}, preserves perceptual order and has the highest discriminative power, allowing users to distinguish and interpret scalar values more accurately. Examples of reconstruction and error patterns for various colormaps are shown in Fig.~\ref{fig:legendExamples}.

These results align well with established principles of colormap design~\cite{bujack2017good}. In particular, we observe that preserving order is the strongest predictor of reconstruction accuracy. Among the colormaps that maintain order, reconstruction error and discriminative power follow a nearly linear trend on a log-log scale, Fig.~\ref{fig:legend}.

Note that these rankings are dataset-dependent. Different datasets would likely produce different orderings, though the general trends and correlations are expected to hold.

\section{3D Colormapping}
When we render a 2D surface in 3D, the visual variable color is used by two different channels. The encoding of the scalar values shares this variable with lighting and shading cues. In this scenario, other colormap quality measures start to play a role, such as redundancy~\cite{rheingans2000task} and robustness to shading on 3D surfaces~\cite{moreland2009diverging}.

\subsection{The Mental Image}
Analogously to 2D colormapping, we assume that observers identify a pixel's color and try to find a match in the legend to interpolate the corresponding value. Humans do not simply go by closest similarity but have an internal color-constancy~\cite{ebner2021color} that allows them to mentally separate the effects of shading from the overall color perception.

\subsection{Reconstruction}
\begin{figure*}[ht]
\subcaptionbox{Absolute difference in hue in CIELCH produced the lowest total $L^2$ error of $3113$. The best performer in this metric is \texttt{Spectral}, the worst is \texttt{cubehelix}.}{\includegraphics[width=0.49\linewidth]{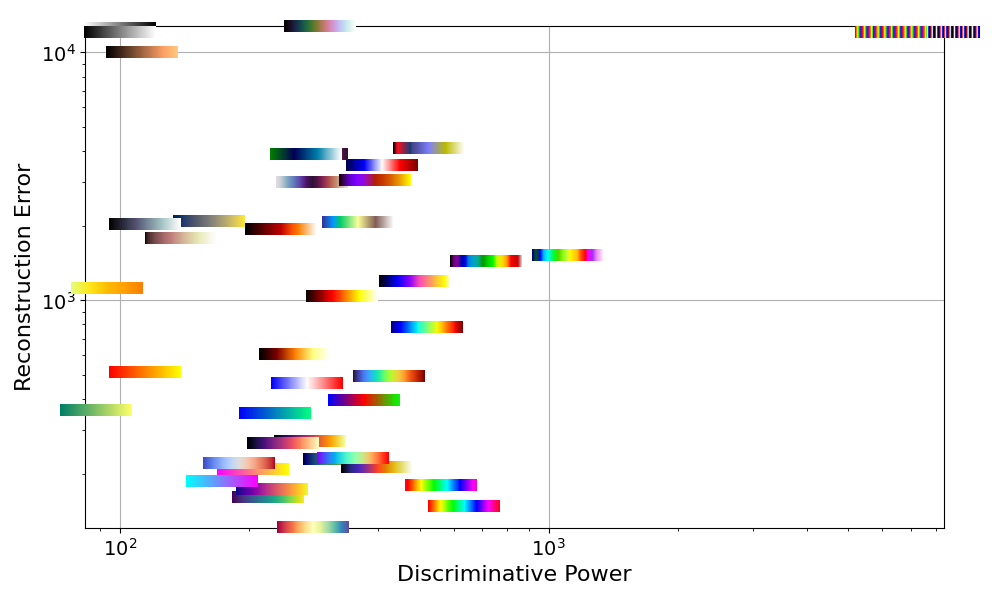}}\hfill
\subcaptionbox{Absolute difference in Euclidean difference in the $ab$ plane in CIELAB produced the second lowest total $L^2$ error of $5092$. The best performer in this metric is \texttt{rainbow}, the worst is \texttt{flag}.}{\includegraphics[width=0.49\linewidth]{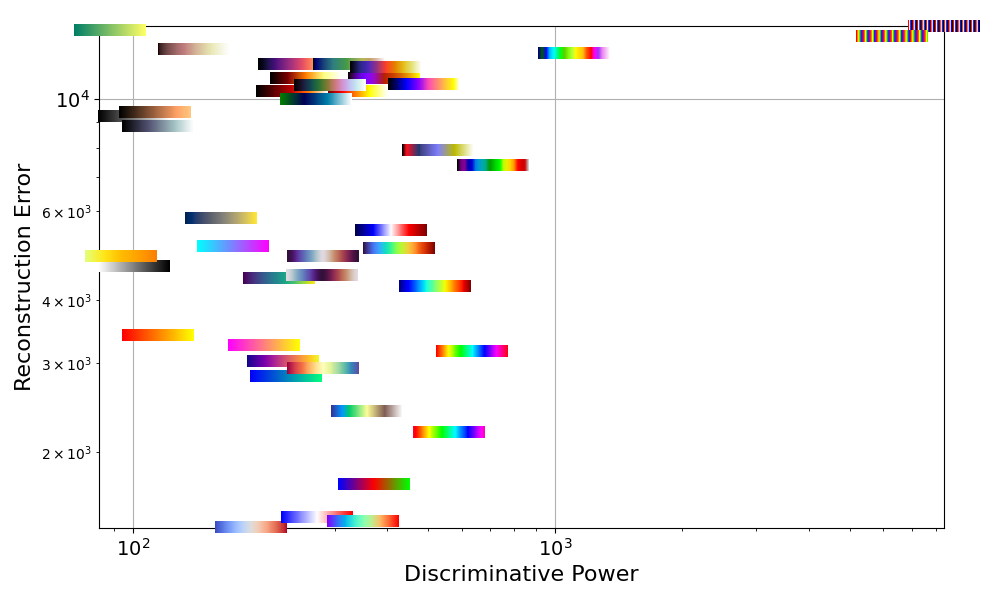}}\hfill
\subcaptionbox{The Euclidean metric $\Delta E_{1976}$ produced a total $L^2$ error of $7098$. The best performer in this metric is \texttt{cool-warm}, the worst is \texttt{flag}.}{\includegraphics[width=0.49\linewidth]{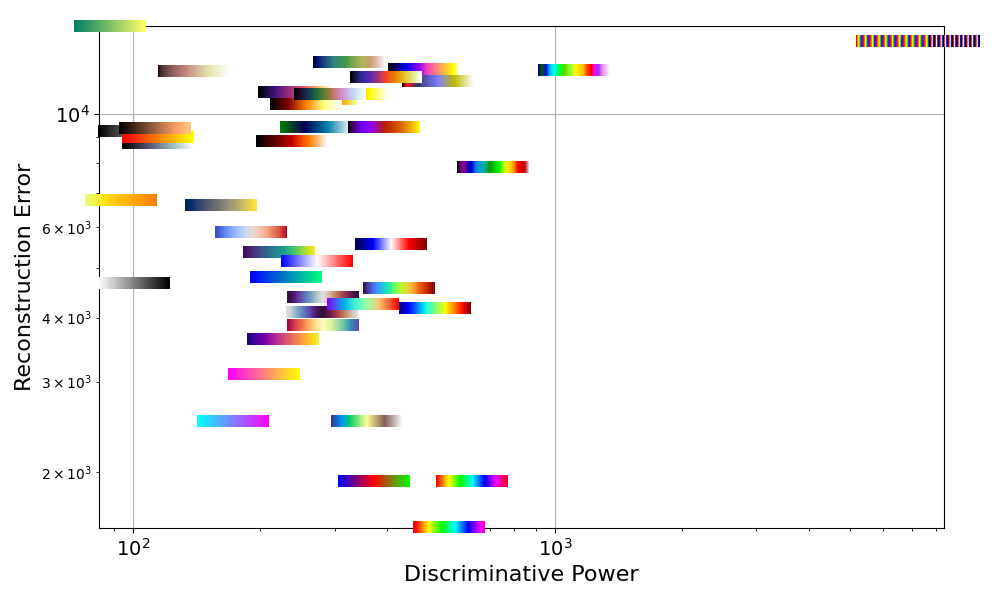}}\hfill
\subcaptionbox{The color distance measure $\Delta E_{2000}$ produced a total $L^2$ error of $7588$. The best performer in this metric is \texttt{gist\_rainbow}, the worst is \texttt{summer}.}{\includegraphics[width=0.49\linewidth]{figs/h/h.png}}
\caption{Log-log plots showing the reconstruction errors of the Matplotlib colormaps for the different metrics.}
\label{fig:3dMetrics}
\end{figure*}

We reconstruct the image analogously to the 2D case, except that we vary the metric for the color lookup to account for color-constancy in a first-order approximative way. We compared four metrics, $\Delta E_{1976}$, $\Delta E_{2000}$, the Euclidean distance in just the $ab$ plane of CIELAB, and the absolute value in just the hue direction of CIELCH. Note that none of these capture the true underlying mechanisms of color-constancy; instead they serve as a first-order approximation for a proof of concept. A comparison of reconstruction errors across all four metrics is shown in Fig.~\ref{fig:3dMetrics}.

\subsection{Evaluation}
\begin{figure}[ht]
    \centering
    \subcaptionbox{\texttt{Spectral}, the best performer in 3D.}{\includegraphics[width=\linewidth]{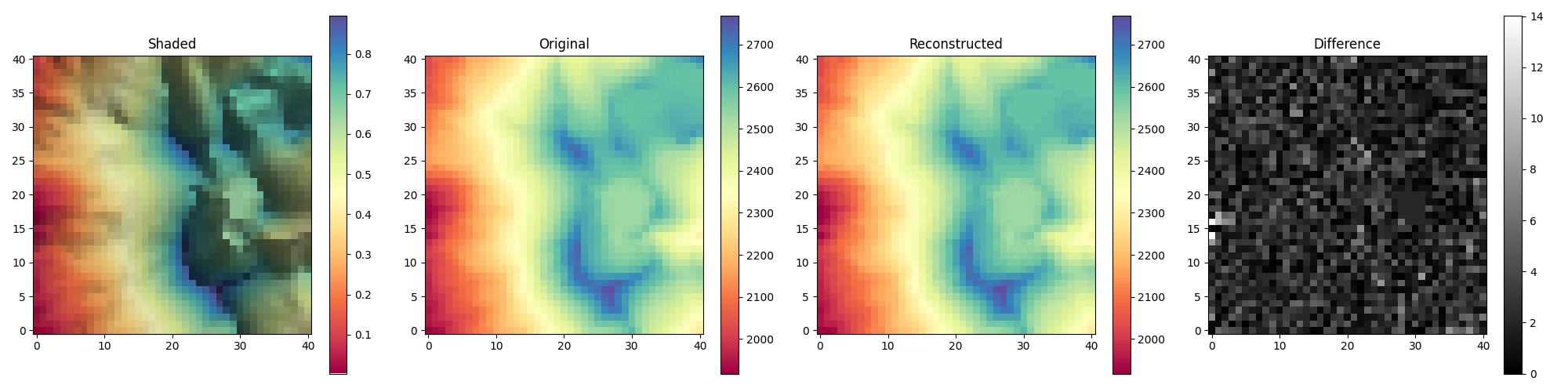}}\hfill
    \subcaptionbox{The perceptually uniform \texttt{gray} cannot reconstruct the shaded image in the hue-based metric.}{\includegraphics[width=\linewidth]{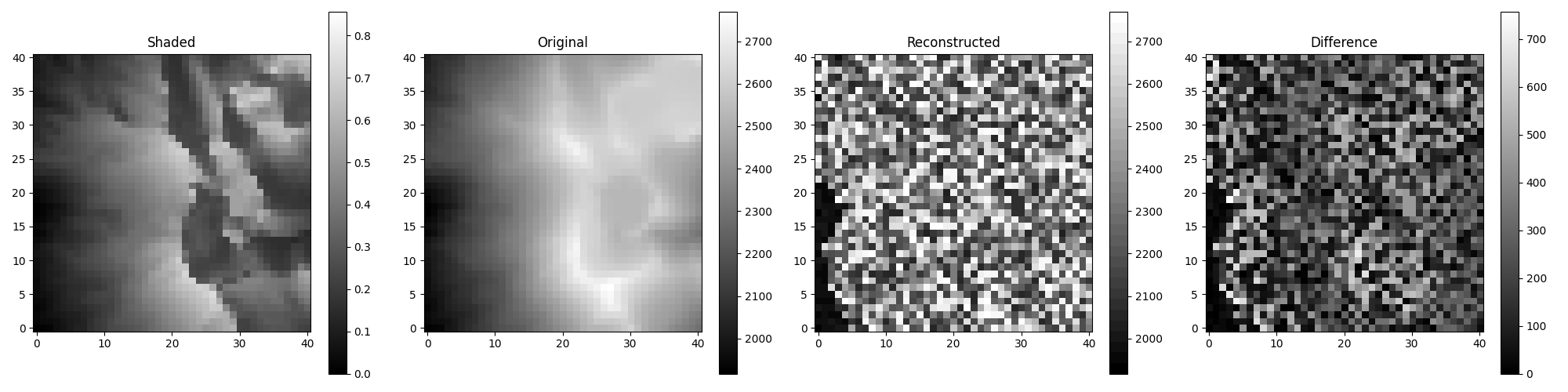}}\hfill
     \subcaptionbox{\texttt{cool-warm} only exhibits problems at the very-red end.}{\includegraphics[width=\linewidth]{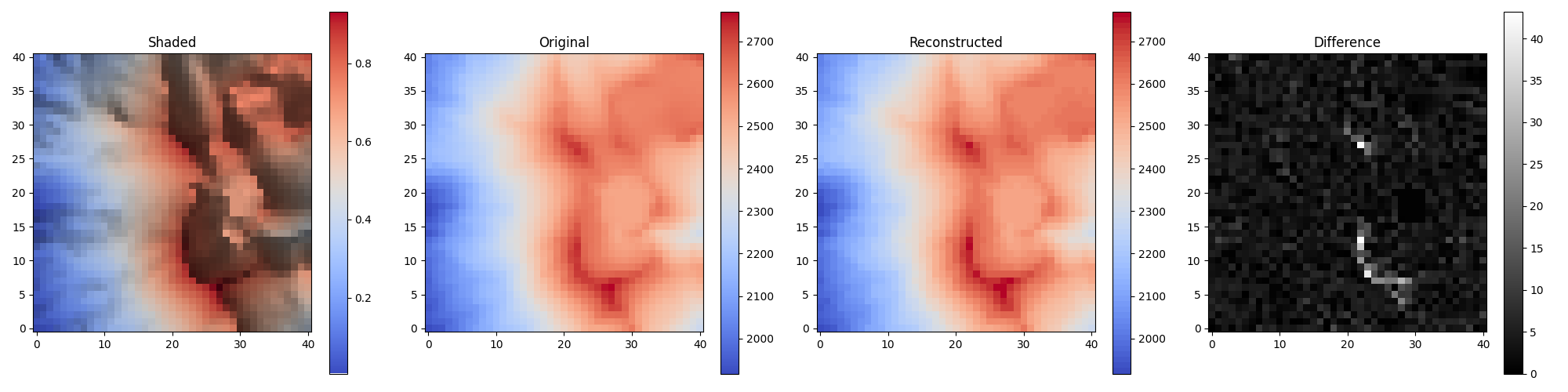}}\hfill
    \subcaptionbox{\texttt{cubehelix} cannot distinguish black from white in the hue-based metric.}{\includegraphics[width=\linewidth]{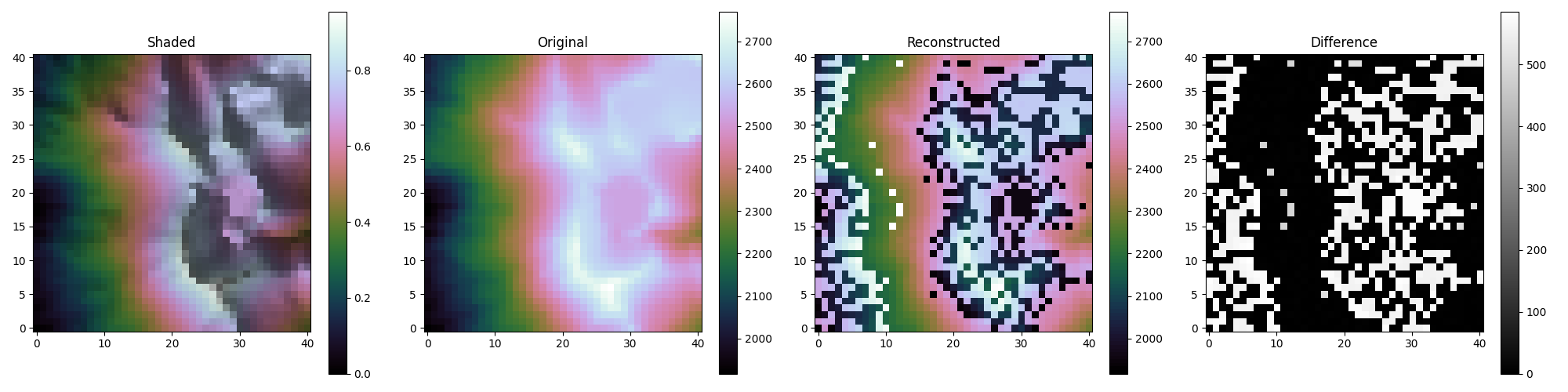}}\hfill
     \subcaptionbox{\texttt{gist\_ncar} has many green shades with very similar hues that cannot be distinguished.}{\includegraphics[width=\linewidth]{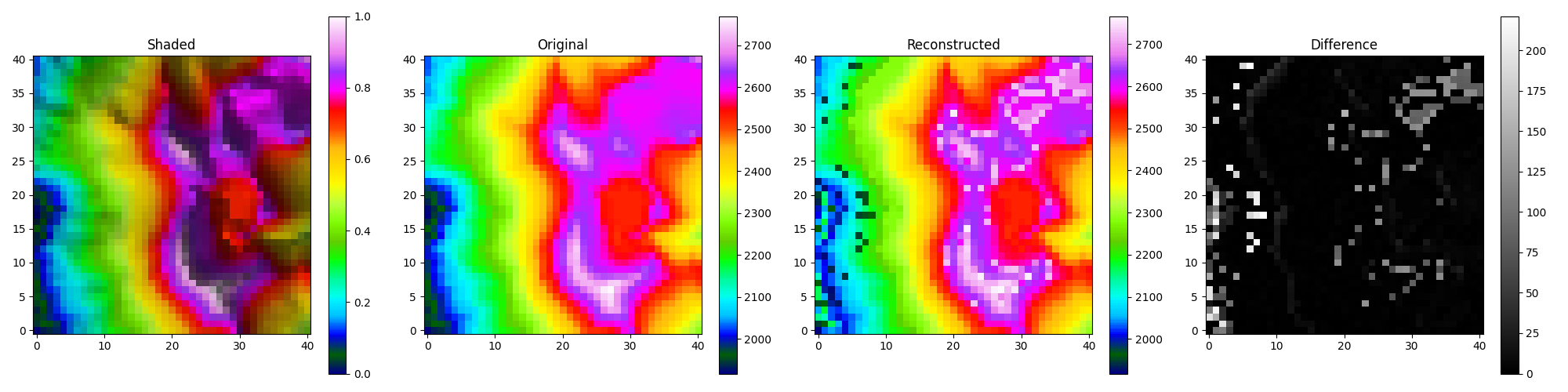}}\hfill
    \caption{Examples of reconstruction and error patterns for the 3D shaded reconstruction task based on the metric using the absolute difference in hue in CIELCH. This metric favors colorful colormaps.}
    \label{fig:3dExamples}
\end{figure}

\begin{figure}[ht]
    \centering
    \includegraphics[width=\linewidth]{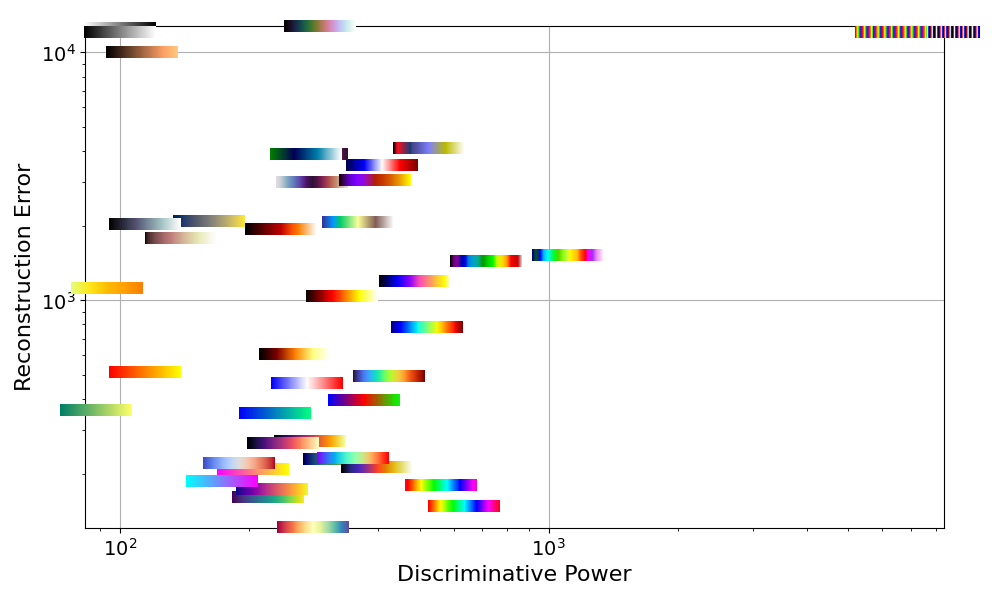}
    \caption{Log-log plot showing the absence of any correlation between the discriminative power and reconstruction error for the reconstruction of shaded images. For the hue-based metric, colorful colormaps produce lower errors because they are more robust to shading in 3D colormapped images.}
    \label{fig:3d}
\end{figure}

We evaluated all Matplotlib colormaps using a subset of PyVista's elevation dataset, restricted to the region $[600, 800] \times [500, 700]$ and subsampled at a $5 \times 5$ resolution. We extruded the dataset by its elevation and applied the corresponding shading when viewing the 3D scene from above.
The best performing metric was just the hue, with the results shown in Fig.~\ref{fig:3d}. The second best performer was the $ab$-plane distance. Examples of reconstruction and error patterns for various colormaps using the hue-based metric are shown in Fig.~\ref{fig:3dExamples}.

\section{Isocontours}
Isocontours are among the most fundamental visualization techniques for 3D scalar fields. They reveal the geometry of regions with equal scalar values and are widely used because of their interpretability and effectiveness.

\subsection{The Mental Image}
When interpreting an isocontour plot, an observer typically references the colorbar to determine the global minimum and maximum, and the specific isovalue, represented by the contour. We provide this same information to our reconstruction algorithm. 

Users intuitively assume that the scalar field takes the given isovalue exactly along the contour. Furthermore, they often assume that the field varies smoothly: values near the contour are similar, while values farther away become increasingly different. If only a single isocontour is provided, there may be an ambiguity if the inside or the outside is the side with higher values. We assume that the observer keeps both options simultaneously in mind. Their natural assumptions form the basis of our reconstruction approach.

\subsection{Reconstruction}
We approximate the original scalar field by computing a signed distance field from the given isocontour. Local extrema of this distance field are identified and scaled to correspond to the global minimum and maximum of the original data. We treat the ambiguity of higher values inside or outside by considering both and selecting the one with the smaller reconstruction error.

To reduce computational cost, we randomly select a subset of $N$ points from the contour, with $N = 500$ in our experiments. These points, along with the global minimum and maximum, are used as interpolation constraints in a radial basis function (RBF) interpolation.
We choose RBFs because they yield smooth scalar fields, especially around extrema. In contrast, linear interpolation can introduce sharp corners, and harmonic interpolation may produce unintuitive cusp-like features. Our approach provides a smooth and plausible scalar field consistent with the isocontour.

\subsection{Evaluation}
To evaluate our method, we generate a synthetic scalar field composed of ten Gaussian kernels with randomly sampled locations, amplitudes, and standard deviations. We compare our approach to the techniques of Kindlmann, Carr, and Bruckner~\etal~\cite{kindlmann1998semi,carr2004simplifying,bruckner2010isosurface}, acknowledging that their methods are not necessarily designed to extract a single isovalue.

\begin{figure}
    \centering
    \subcaptionbox{Kindlmann.}{\includegraphics[width=0.49\linewidth]{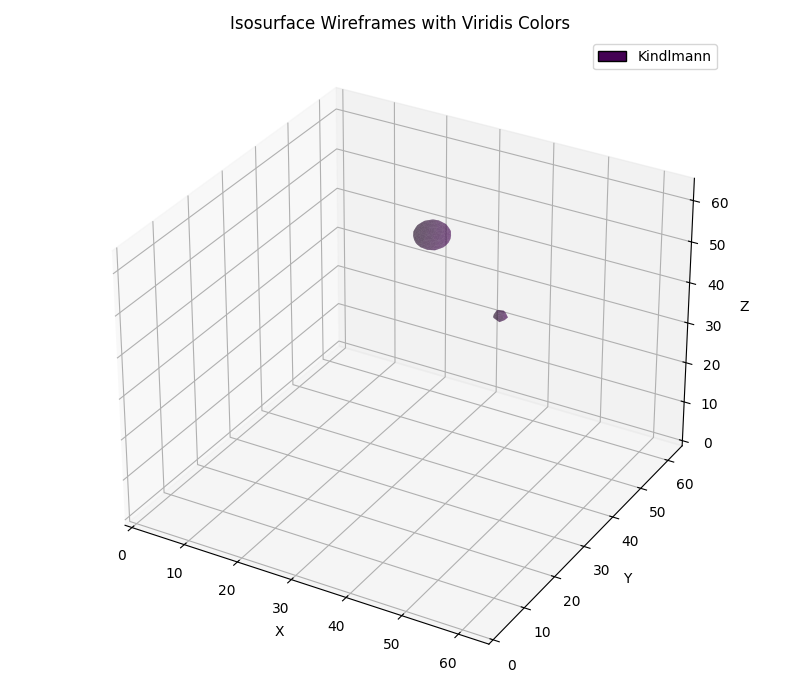}}\hfill
    \subcaptionbox{Bruckner.}{\includegraphics[width=0.49\linewidth]{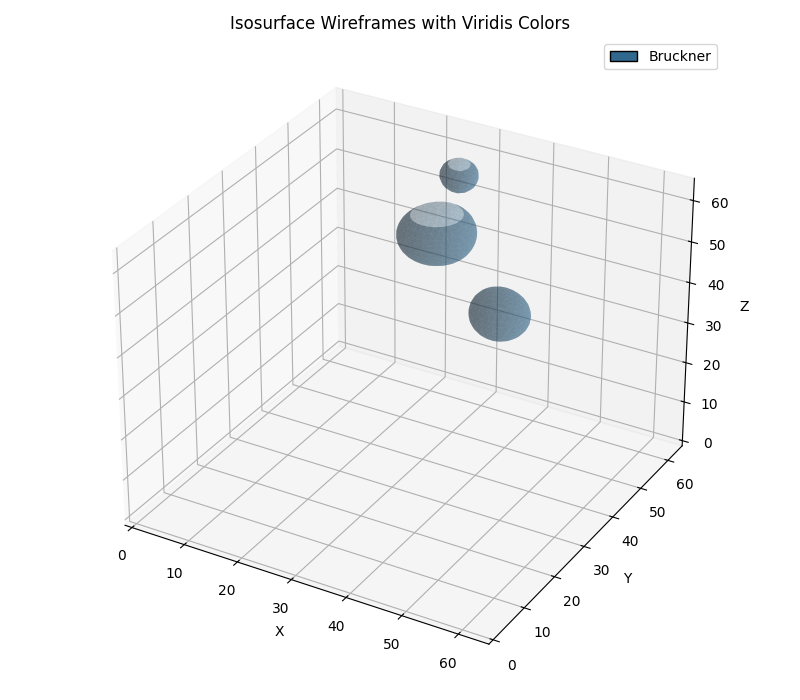}}\hfill
    \subcaptionbox{Carr.}{\includegraphics[width=0.49\linewidth]{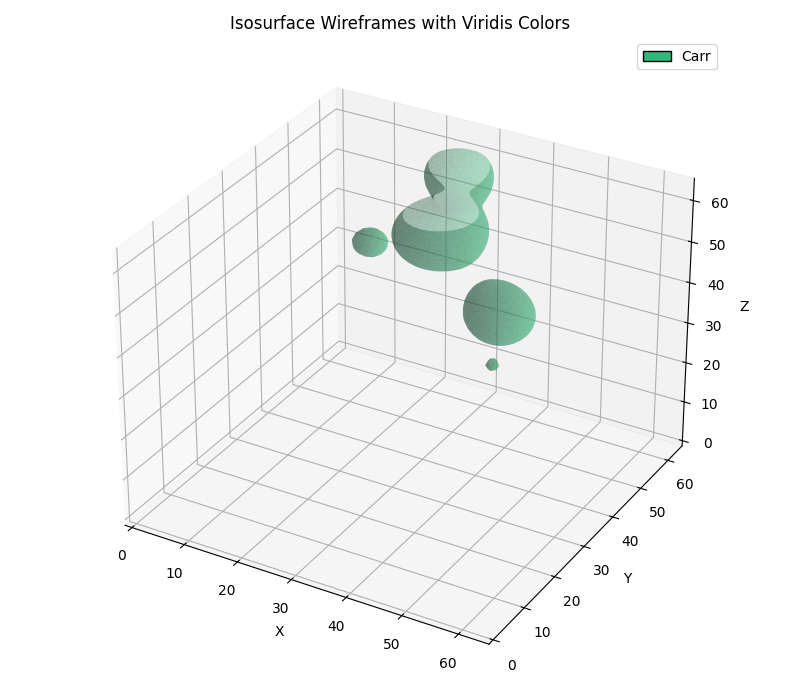}}\hfill
    \subcaptionbox{Ours.}{\includegraphics[width=0.49\linewidth]{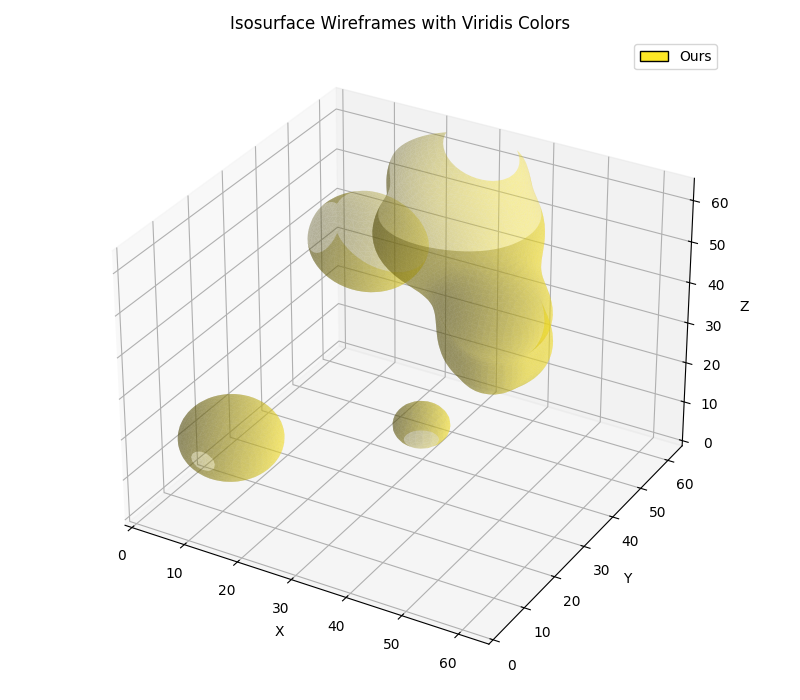}}\hfill
    \caption{Contours chosen by different methods. Our reconstruction-error-based method provides a better balance between covering the domain and representing the scalar value range.}
    \label{fig:contour}
\end{figure}

In this context, Kindlmann's method effectively selects the isovalue at the location of maximum gradient magnitude, while Carr's method reduces to using the midpoint of the data range. Bruckner's approach chooses an isovalue based on global isosurface similarity. All three methods tend to select relatively high isovalues, resulting in contours that enclose only a small portion of the domain. These choices highlight regions near maxima but do not provide a strong mental image of the overall structure of the data, Fig.~\ref{fig:contour}.

In contrast, our data-reconstruction metric yields an isovalue that offers a better trade-off between spatial domain coverage and scalar value representation. The example shown in Fig.~\ref{fig:contour} is representative of the typical behavior of each method across other randomly generated Gaussian datasets.

\section{Viewing Angle}
Viewing angle strongly influences how well a 3D visualization conveys the structure of the underlying data. Different viewpoints expose or hide critical features, and users build their mental image of the scene from the visible surfaces alone. In our framework, we treat the viewing angle as another visualization parameter and evaluate it through reconstruction quality, allowing us to automatically identify viewpoints that best support an accurate mental model.
\subsection{The Mental Image} We assume that an observer, by slightly moving their head or using a stereoscopic view, forms an approximate impression of the depth of the object's side facing the camera. The silhouette of the object is clearly perceived, while the depth of the side facing away remains largely unknown.
\subsection{Reconstruction} NeRF reconstructs the density field using nine images: the central view and eight neighboring views at offsets of $\pm 10^\circ$. This yields both a density field and an RGB field. The object's surface is then approximated by applying an isocontouring technique at the median density value.
\subsection{Evaluation} 
\begin{figure}[ht]
    \subcaptionbox{The Hausdorff distance (lower is better) between ground-truth teapot reconstruction prefers a slightly elevated side view. We filtered out the 10\% worst reconstructions that correspond to outliers where the NeRF did not converge.}{
        \includegraphics[width=\linewidth]{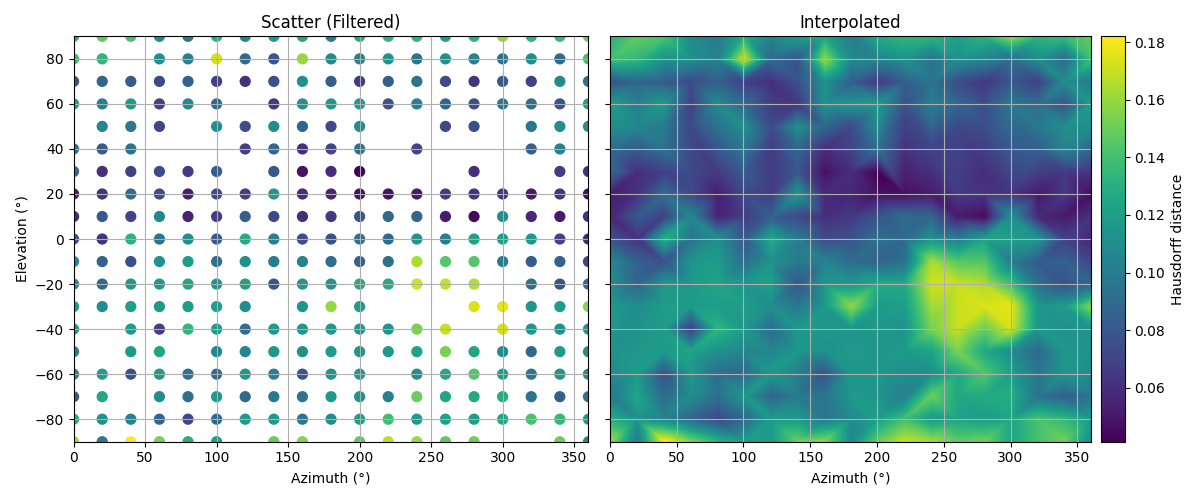}
    }\hfill
    \subcaptionbox{The chamfer distance (lower is better) between ground-truth teapot reconstruction prefers a slightly elevated side view. We filtered out the 10\% worst reconstructions that correspond to outliers where the NeRF did not converge.}{
        \includegraphics[width=\linewidth]{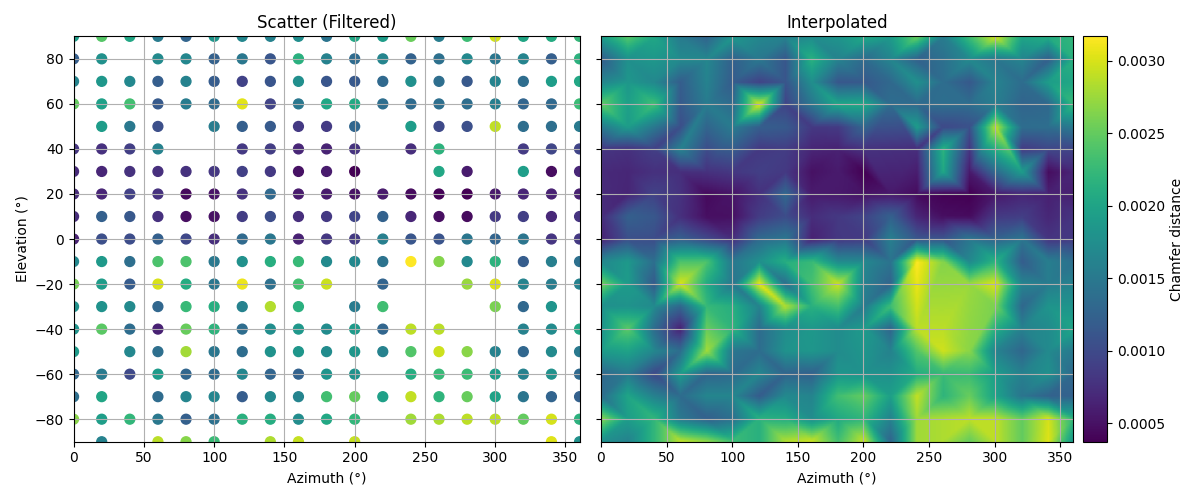}
    }\hfill
    \subcaptionbox{Entropy (higher is better) prefers top-down or bottom-up views.}{
        \includegraphics[width=\linewidth]{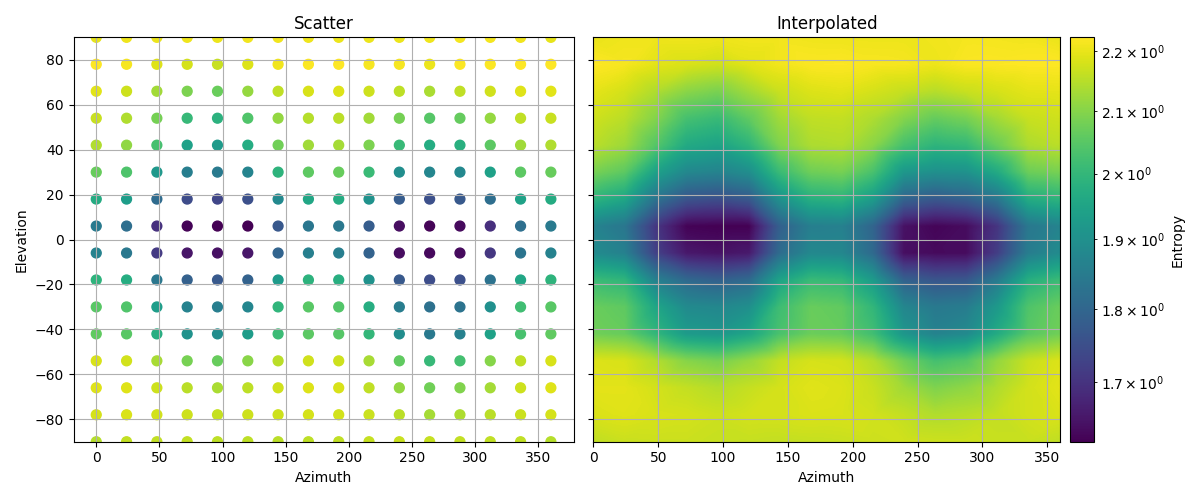}
    }
   
    \caption{The preferred orientations from reconstruction error and entropy presented as points and linearly interpolated.}
    \label{fig:teapot}
\end{figure}

\begin{figure}[ht]\captionsetup[subfigure]{labelformat=empty}
    \subcaptionbox{}{
\includegraphics[width=0.17\linewidth, trim=5cm 3.8cm 5cm 3.8cm, clip]{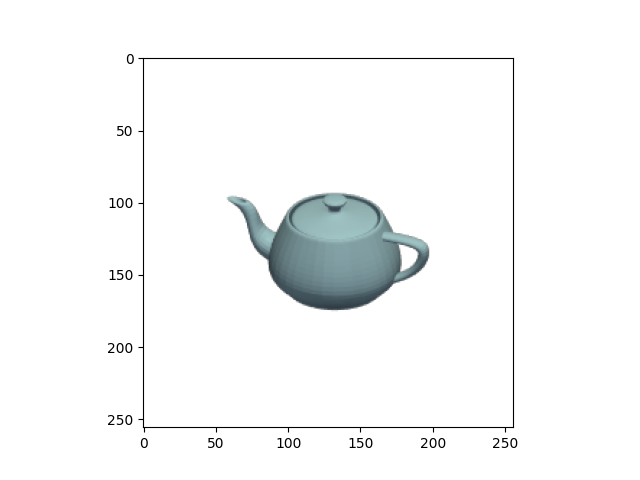}
    }\hfill
    \subcaptionbox{}{
\includegraphics[width=0.17\linewidth, trim=5cm 3.8cm 5cm 3.8cm, clip]{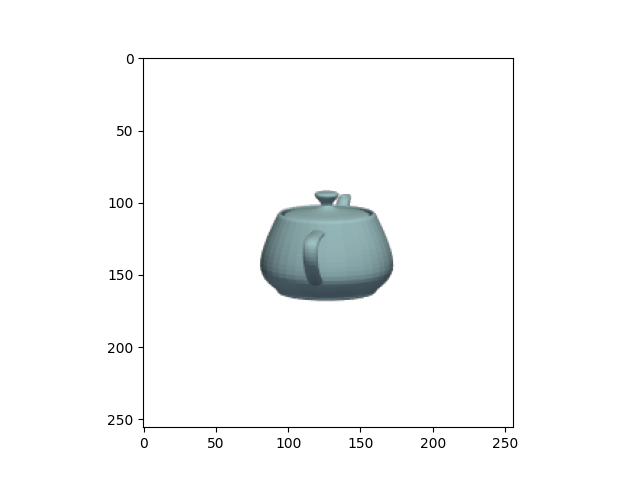}
    }\hfill
    \subcaptionbox{}{
\includegraphics[width=0.17\linewidth, trim=5cm 3.8cm 5cm 3.8cm, clip]{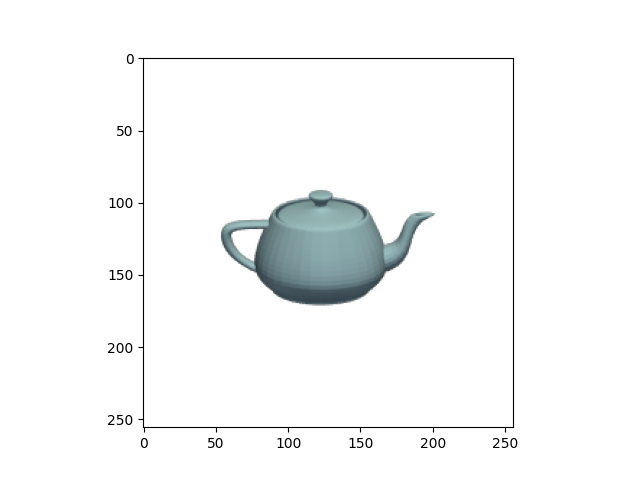}
    }\hfill
    \subcaptionbox{}{
\includegraphics[width=0.17\linewidth, trim=5cm 3.8cm 5cm 3.8cm, clip]{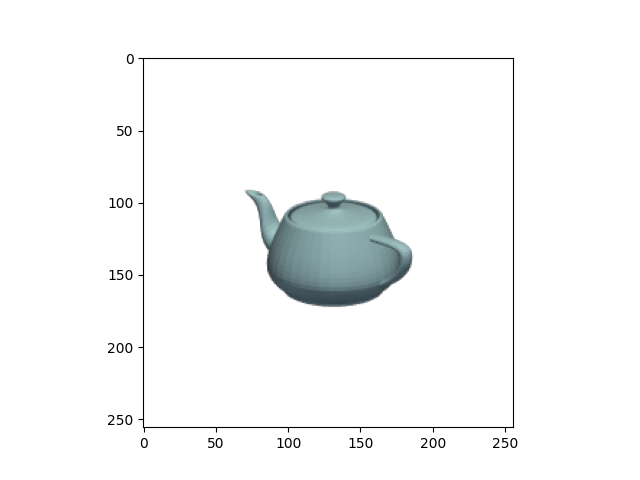}
    }\hfill
    \subcaptionbox{}{
\includegraphics[width=0.17\linewidth, trim=5cm 3.8cm 5cm 3.8cm, clip]{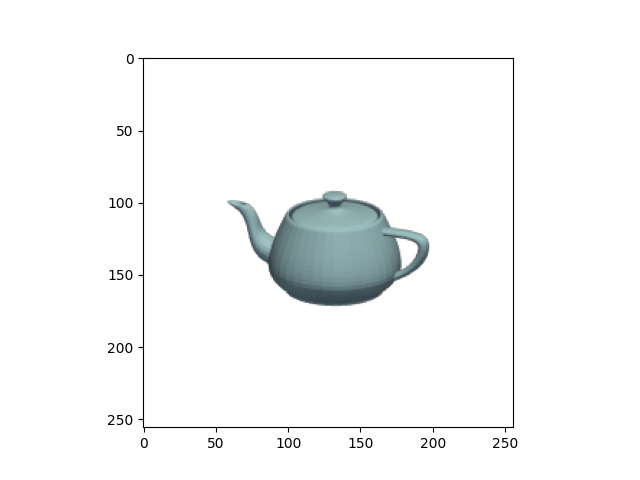}
    }\hfill
    \subcaptionbox{}{
\includegraphics[width=0.17\linewidth, trim=5cm 3.8cm 5cm 3.8cm, clip]{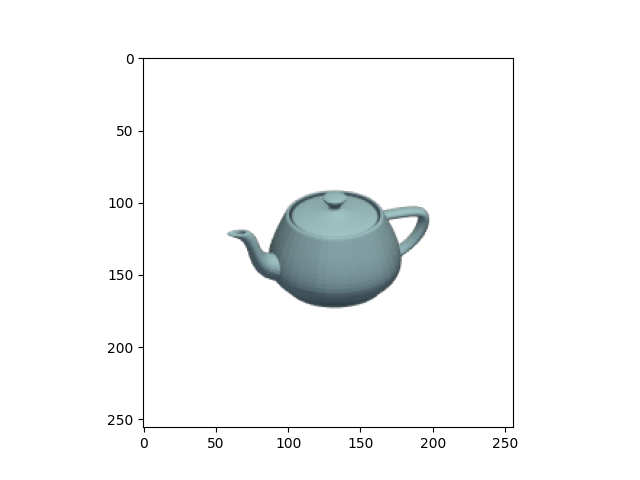}
    }\hfill
    \subcaptionbox{}{
\includegraphics[width=0.17\linewidth, trim=5cm 3.8cm 5cm 3.8cm, clip]{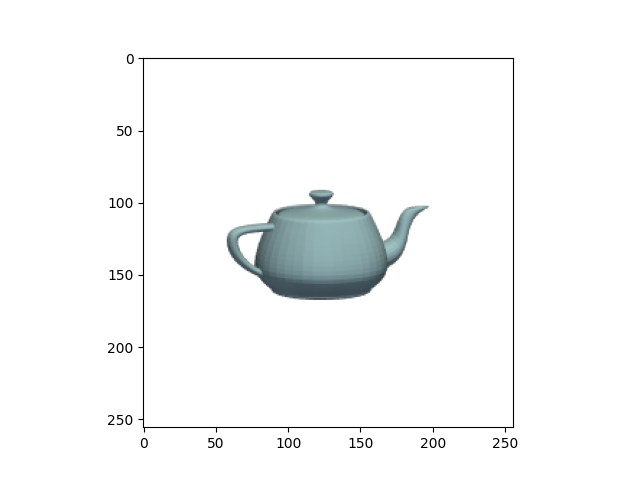}
    }\hfill
    \subcaptionbox{}{
\includegraphics[width=0.17\linewidth, trim=5cm 3.8cm 5cm 3.8cm, clip]{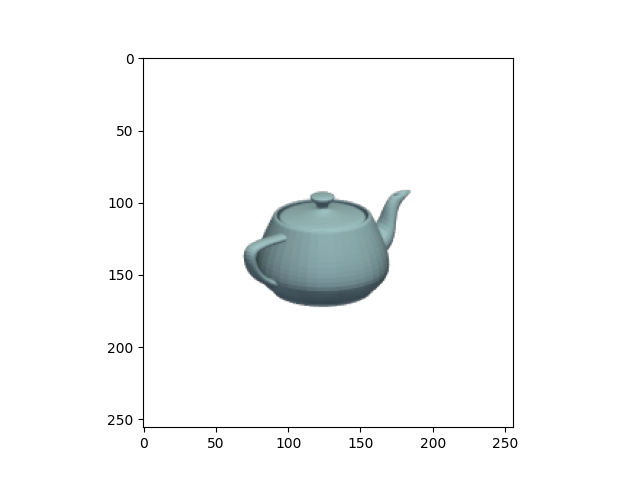}
    }\hfill
    \subcaptionbox{}{
\includegraphics[width=0.17\linewidth, trim=5cm 3.8cm 5cm 3.8cm, clip]{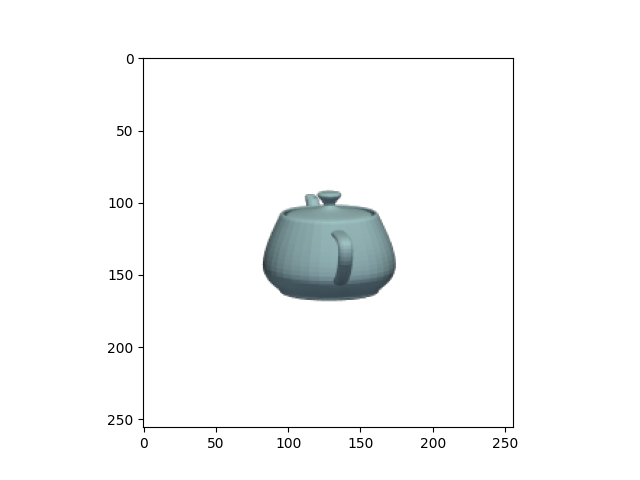}
    }\hfill
    \subcaptionbox{}{
\includegraphics[width=0.17\linewidth, trim=5cm 3.8cm 5cm 3.8cm, clip]{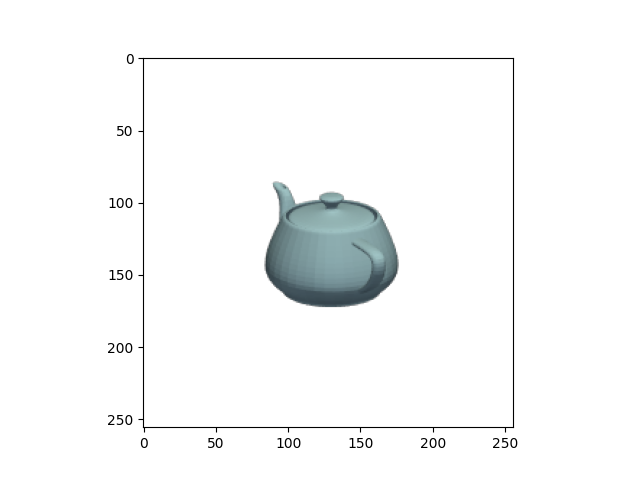}
    }\hfill
    \caption{The 10 best performing orientations of the Utah teapot with respect to reconstruction accuracy are mostly side views.}
    \label{fig:teapotGood}
\end{figure}

\begin{figure}[ht]\captionsetup[subfigure]{labelformat=empty}
    \subcaptionbox{}{
\includegraphics[width=0.17\linewidth, trim=5cm 3.8cm 5cm 3.8cm, clip]{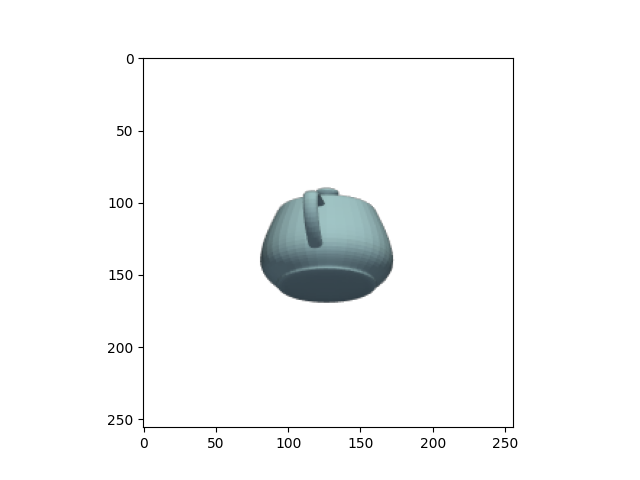}
    }\hfill
    \subcaptionbox{}{
\includegraphics[width=0.17\linewidth, trim=5cm 3.8cm 5cm 3.8cm, clip]{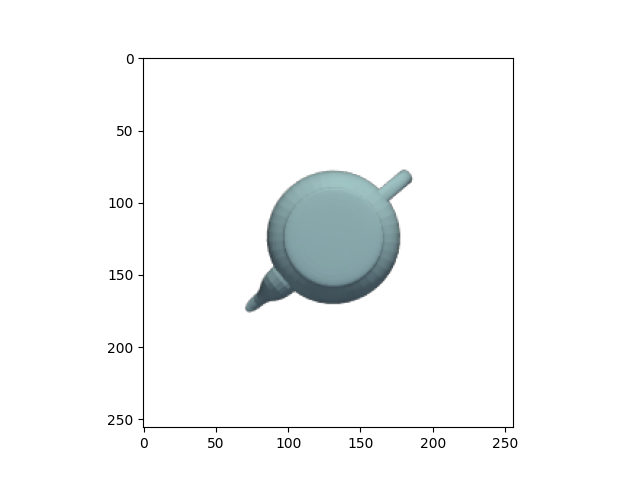}
    }\hfill
    \subcaptionbox{}{
\includegraphics[width=0.17\linewidth, trim=5cm 3.8cm 5cm 3.8cm, clip]{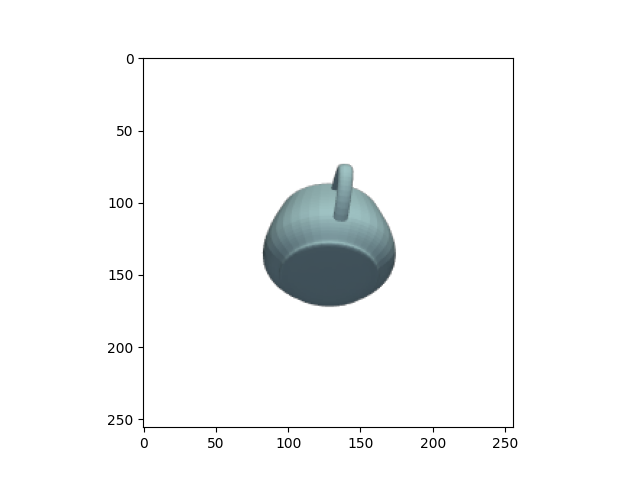}
    }\hfill
    \subcaptionbox{}{
\includegraphics[width=0.17\linewidth, trim=5cm 3.8cm 5cm 3.8cm, clip]{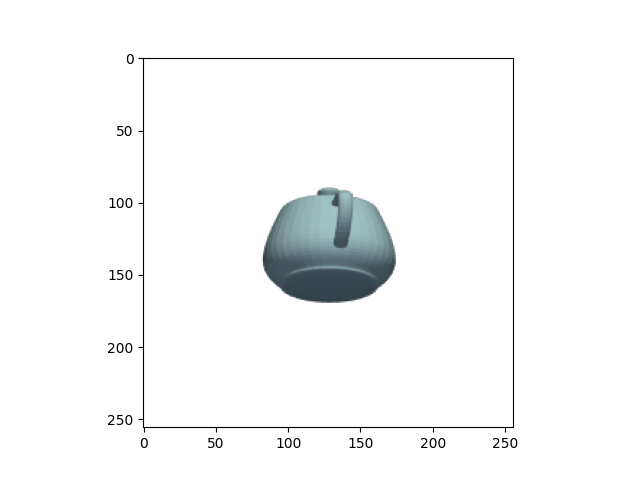}
    }\hfill
    \subcaptionbox{}{
\includegraphics[width=0.17\linewidth, trim=5cm 3.8cm 5cm 3.8cm, clip]{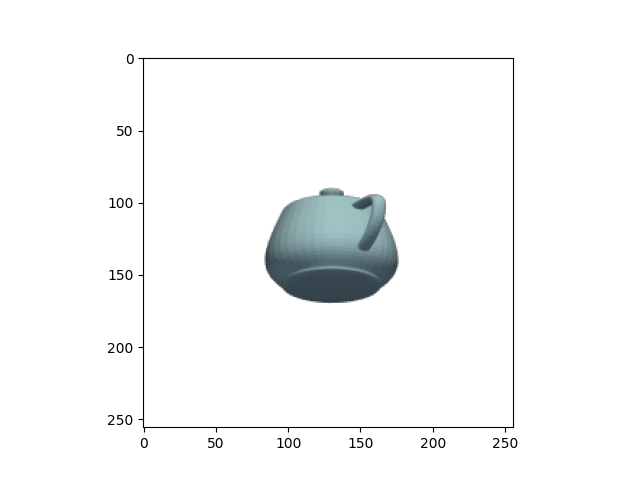}
    }\hfill
    \subcaptionbox{}{
\includegraphics[width=0.17\linewidth, trim=5cm 3.8cm 5cm 3.8cm, clip]{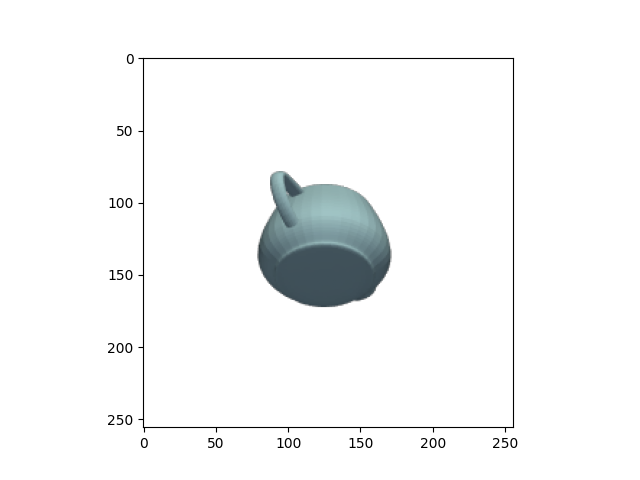}
    }\hfill
    \subcaptionbox{}{
\includegraphics[width=0.17\linewidth, trim=5cm 3.8cm 5cm 3.8cm, clip]{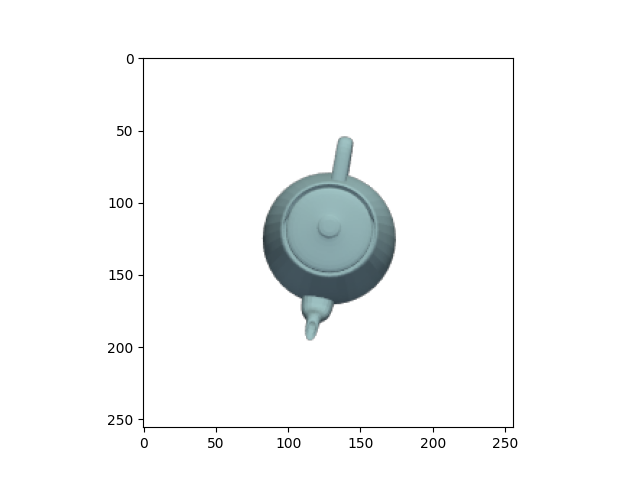}
    }\hfill
    \subcaptionbox{}{
\includegraphics[width=0.17\linewidth, trim=5cm 3.8cm 5cm 3.8cm, clip]{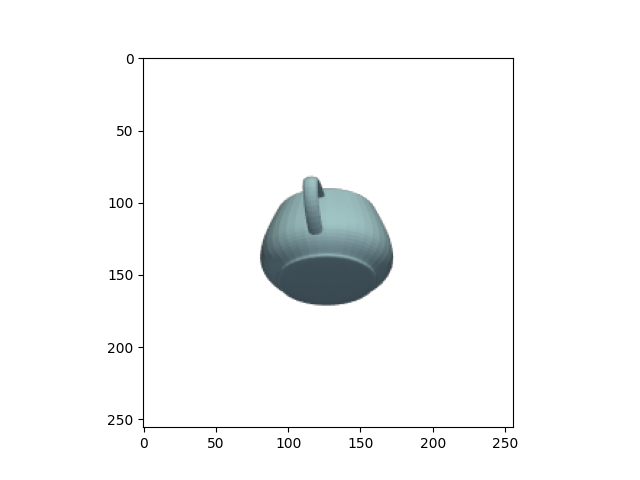}
    }\hfill
    \subcaptionbox{}{
\includegraphics[width=0.17\linewidth, trim=5cm 3.8cm 5cm 3.8cm, clip]{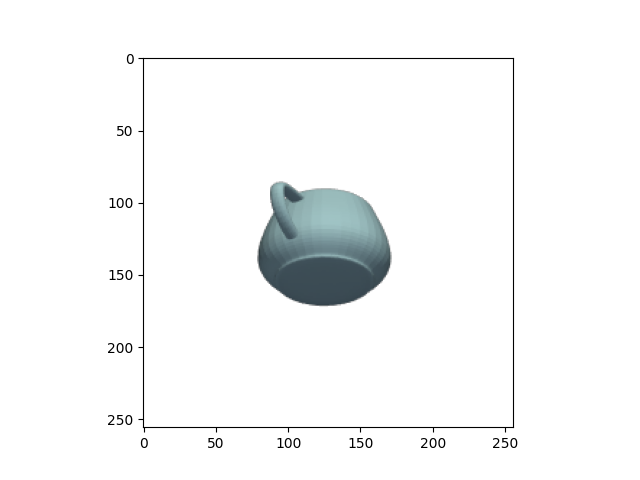}
    }\hfill
    \subcaptionbox{}{
\includegraphics[width=0.17\linewidth, trim=5cm 3.8cm 5cm 3.8cm, clip]{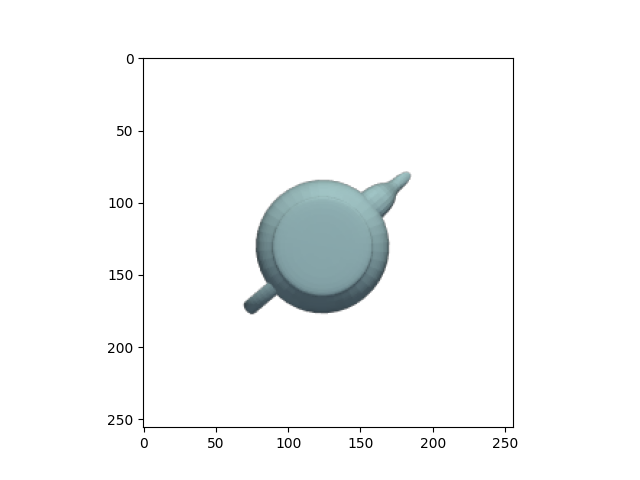}
    }\hfill
    \caption{The 10 worst-performing orientations of the Utah teapot with respect to reconstruction accuracy are mostly angles where the spout and/or the lid are hidden. Some angles from top-down or bottom-up also perform badly.}
    \label{fig:teapotBad}
\end{figure}

\begin{figure}[ht]
        \includegraphics[width=\linewidth]{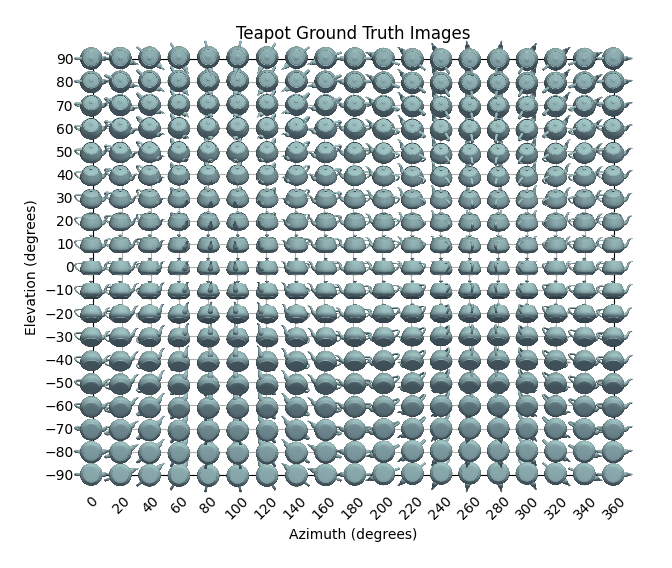}
        \caption{Ground truth renderings of the teapot at the central view.}
    \label{fig:teapotLarge}
\end{figure}

We compute the Hausdorff and chamfer distances between the original surface and the reconstructed surface of the Utah teapot, as shown in Fig.~\ref{fig:teapot}. With the help of the legend of orientations in Fig.~\ref{fig:teapotLarge}, the results show a clear preference for views from slightly above ($20^\circ$ elevation) and a small preference against views from the back ($270^\circ$ azimuth). The 10 best orientations are shown in Fig.~\ref{fig:teapotGood}. They all share the slightly downward perspective that shows the lid well. The worst orientations are from mostly the bottom, where the lid cannot be seen or from the back where the spout cannot be seen, Fig.~\ref{fig:teapotBad}. 

We compare our results to the idea proposed by Vázquez et al.\ of favoring high-entropy images over low-entropy ones~\cite{vazquez2001viewpoint}, Fig.~\ref{fig:teapot}(c). Entropy suggests almost the opposite: top-down and bottom-up views appear preferable, likely because the teapot’s widest extent is most visible there. This is counterintuitive because these views do not reveal critical structures such as the hole in the handle.

\begin{figure}[ht]
    \subcaptionbox{Ground Truth angles 0, 0}{
\includegraphics[height=0.29\linewidth]{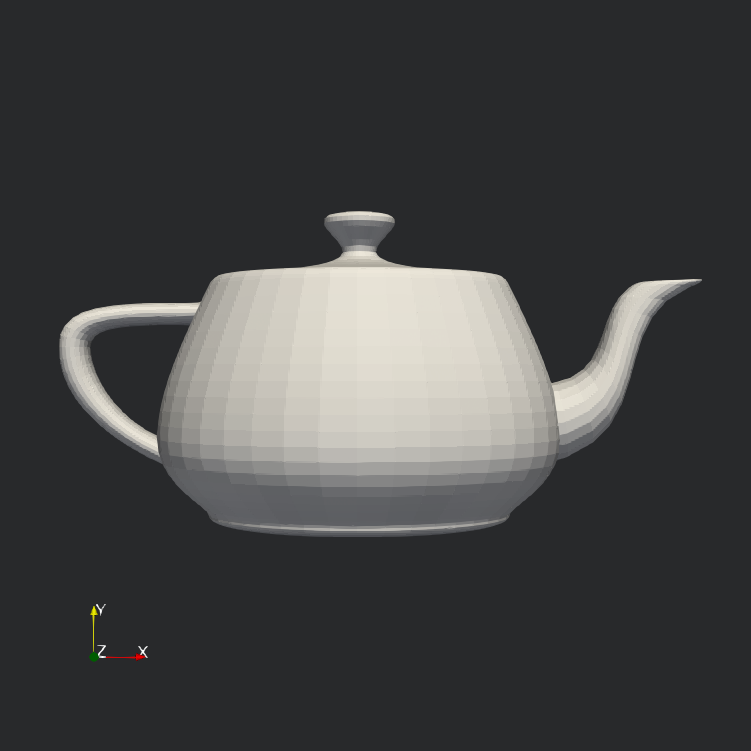}
    }\hfill
    \subcaptionbox{Good reconstruction around 0, 0 viewed from 0, 0}{
        \includegraphics[height=0.29\linewidth]{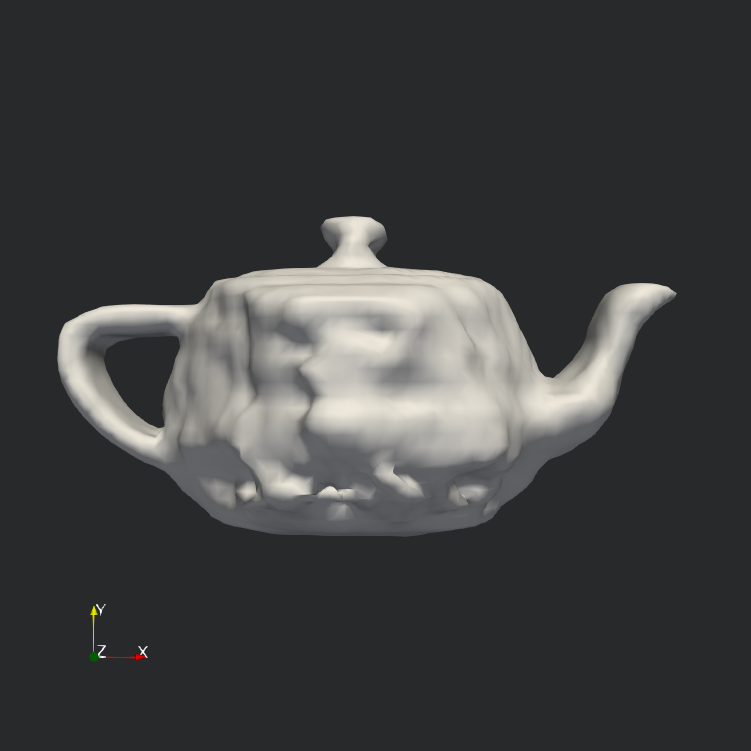}
    }\hfill
    \subcaptionbox{Bad reconstruction around 270, 0 viewed from 0, 0}{
        \includegraphics[height=0.29\linewidth]{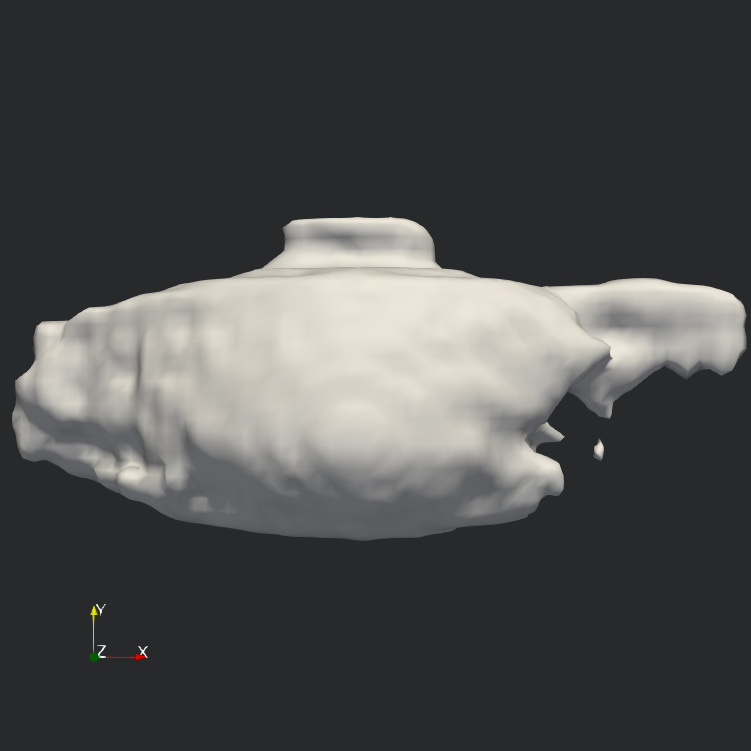}
    }\hfill
    \subcaptionbox{Ground Truth angles 270, 0}{
        \includegraphics[height=0.29\linewidth]{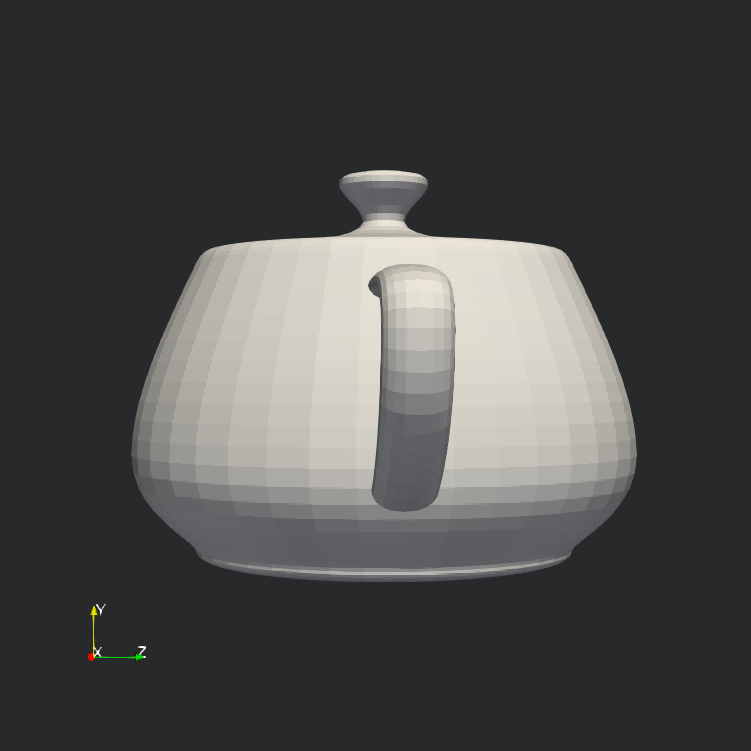}
    }\hfill
    \subcaptionbox{Good reconstruction around 0, 0 viewed from 270, 0}{
        \includegraphics[height=0.29\linewidth]{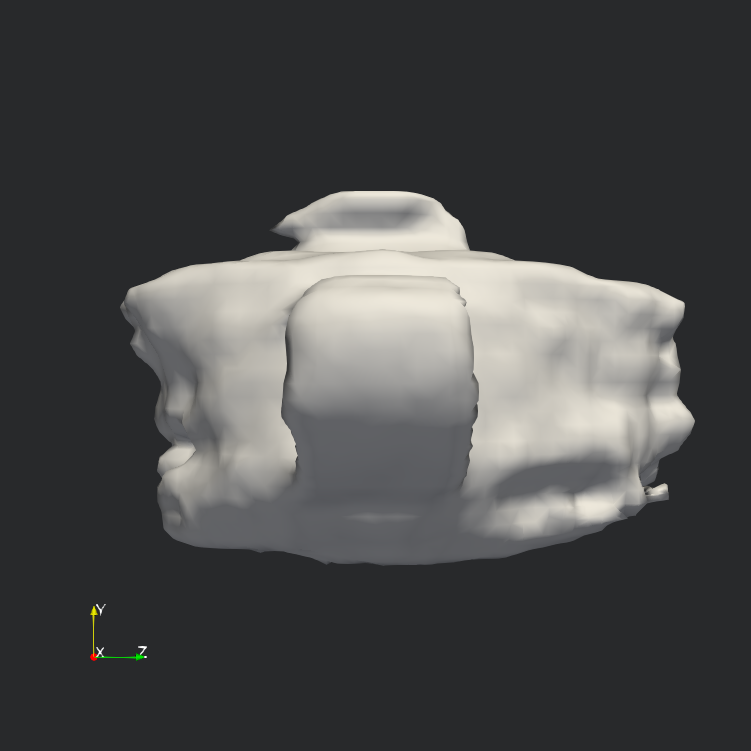}
    }\hfill
    \subcaptionbox{Bad reconstruction around 270, 0 viewed from 270, 0}{
        \includegraphics[height=0.29\linewidth]{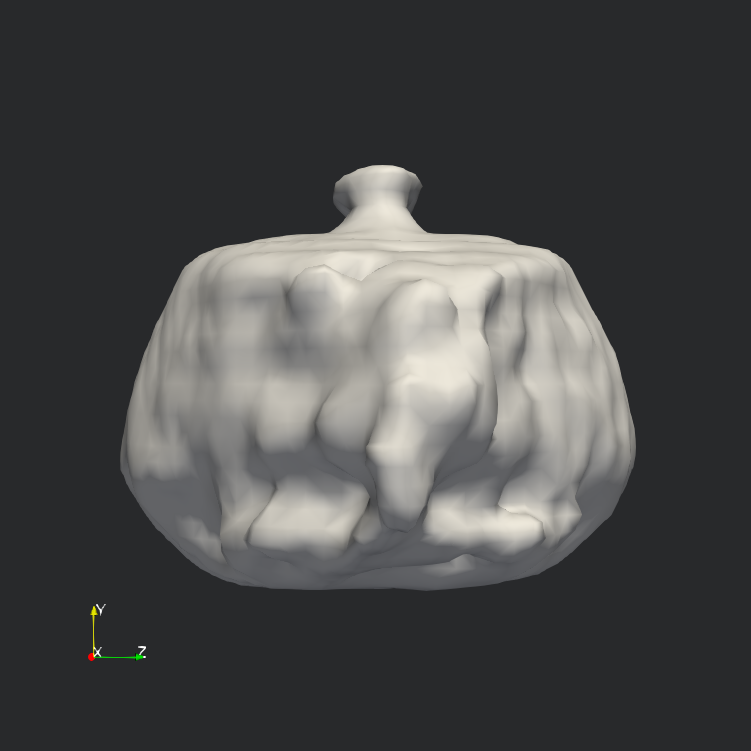}
    }\hfill
    \caption{Examples of reconstruction based on NeRF for the Utah teapot. The side facing the camera is captured very well. The side orthogonal to the camera typically exhibits considerable uncertainty. Some orientations are superior to others, for example, the view from the back does not know anything about the teapot's spout, whereas the view from the side has fewer hidden features.}
    \label{fig:teapotRec}
\end{figure}

An intuition about why some viewing angles perform better than others can be gained from examining Fig.~\ref{fig:teapotRec}. It can be seen that the reconstruction of a basic NeRF is inferior to humans' mental images who take into account symmetries and similar objects they have seen in the past from all sides. Again, this paper is a proof of concept and improving the reconstruction by adding prior knowledge, e.g., as in diffusion models~\cite{gao2024cat3d}, will automatically improve the quality of the metric.




\section{Full Visualization Workflow}
We combine the previous examples into a full workflow, where we look for a good colormap, isovalue, and camera angle simultaneously. It is possible to use traditional metrics for each of these tasks and choose the best parameter for each individually, but the fact that the metrics are all valid specifically for one part of the overall workflow means that we cannot take into account cross-influence. Our suggested metric, on the other hand, applies to every part of a visualization workflow and can therefore be used to evaluate good parameters while taking cross-effects into account. For example, there might be an isovalue that covers the field well because it appears in a nested or layered way multiple times but the inner occurrences do not contribute to a better understanding because the user does not see them from the outside. This type of behavior cannot be captured by evaluating the components individually but it can be identified by our metric because of its universal applicability.


\subsection{The Mental Image}
We assume that observers combine the mental images from the previous sections. They estimate the shape of the surface and look up the color in the colorbar to identify the scalar value at the location of the surface and the overall minimum and maximum. They then assume the scalar value to deviate more from the looked-up value the further the location is away from the surface.

\subsection{Reconstruction}
We run NeRF and extract the approximate isocontour by computing the contour of the smoothed density field at its mean value.
We take the RGB field and cluster it in CIELAB ($\Delta E_{2000}$) into two clusters using k-means: one corresponding to unknown values in transparent regions (close to black), and the other to the contour color.
We select the color center farther from black and match it to the closest color in the color legend using only the hue to reduce shading interference.
We then approximate the 3D scalar field by interpolating the signed distance field using the identified scalar value.

\subsection{Evaluation}
\begin{figure}[ht]
\captionsetup[subfigure]{labelformat=empty}
\subcaptionbox{}{
        \includegraphics[width=0.47\linewidth]{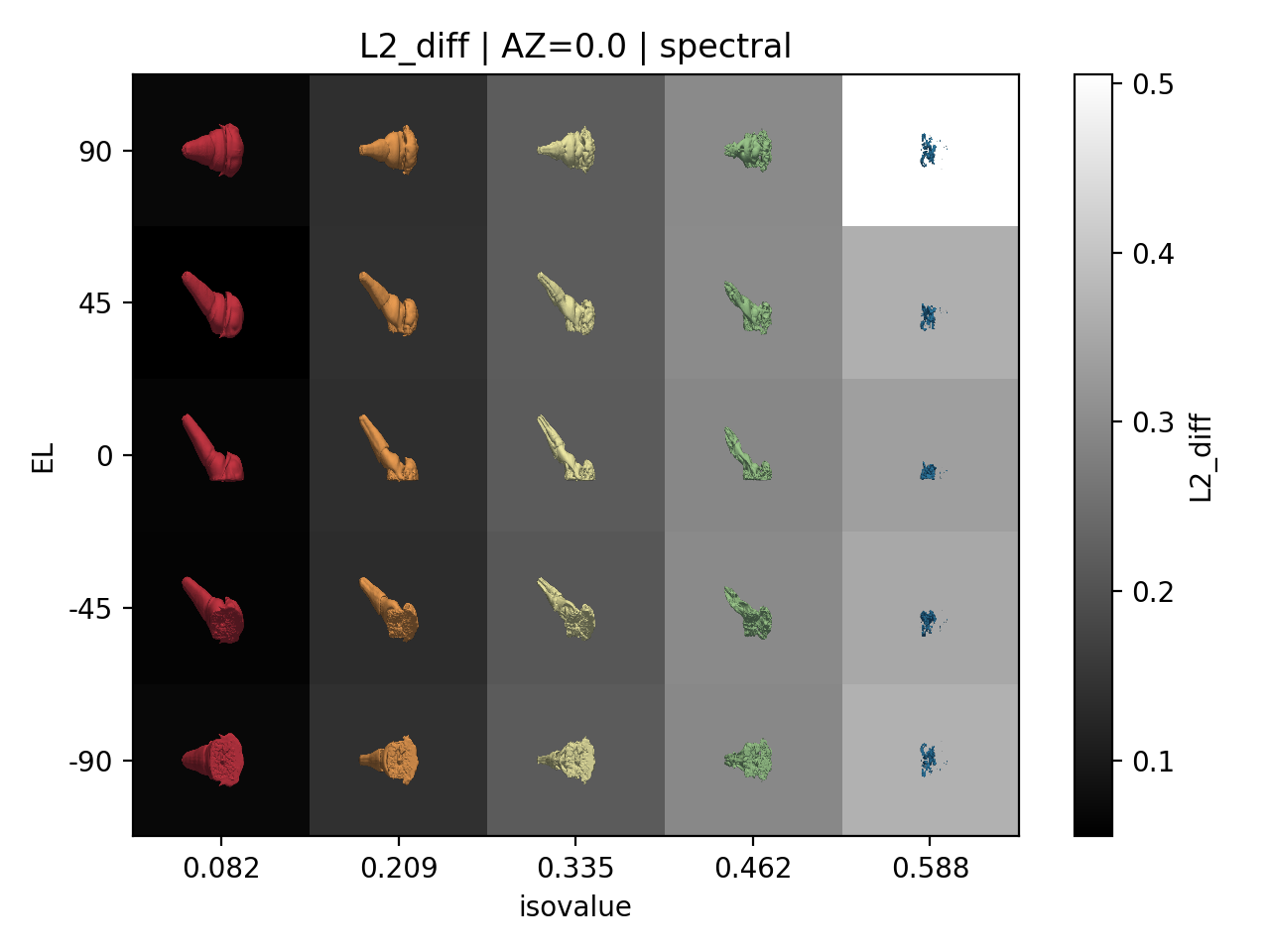}
    }\hfill
\subcaptionbox{}{
        \includegraphics[width=0.47\linewidth]{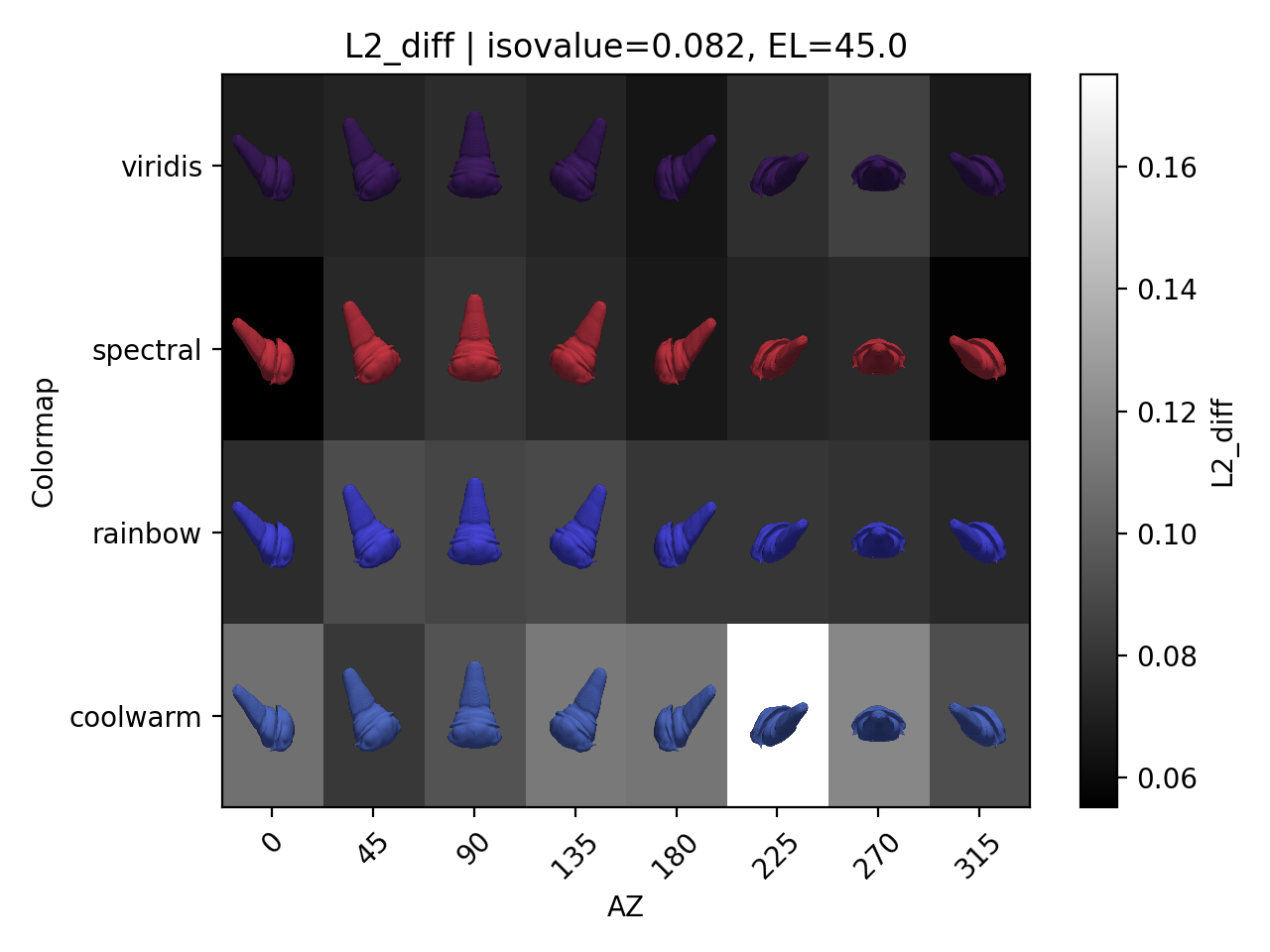}
    }\hfill
\subcaptionbox{}{
        \includegraphics[width=0.47\linewidth]{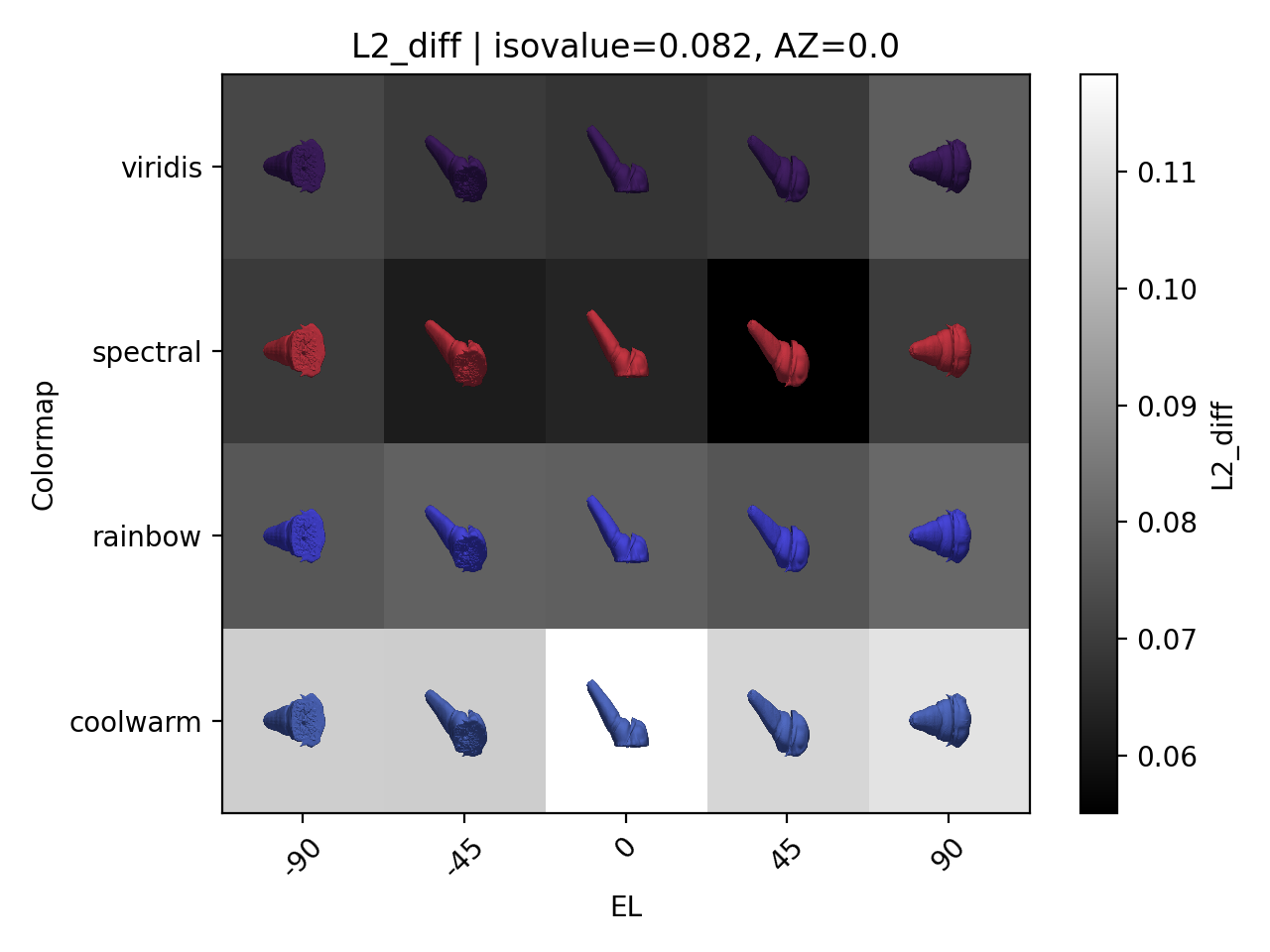}
    }\hfill
\subcaptionbox{}{
        \includegraphics[width=0.47\linewidth]{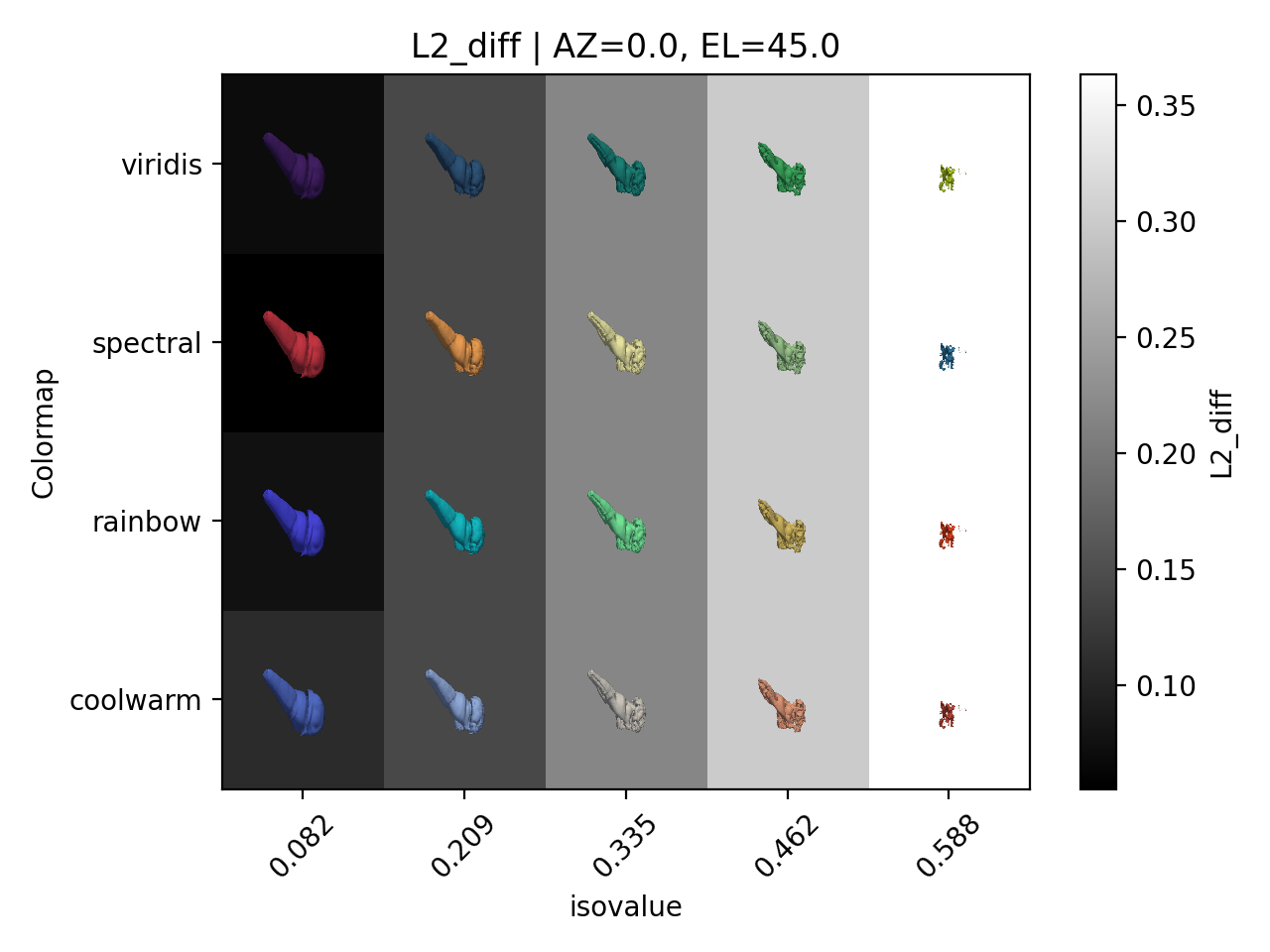}
    }\hfill
\subcaptionbox{}{
        \includegraphics[width=0.47\linewidth]{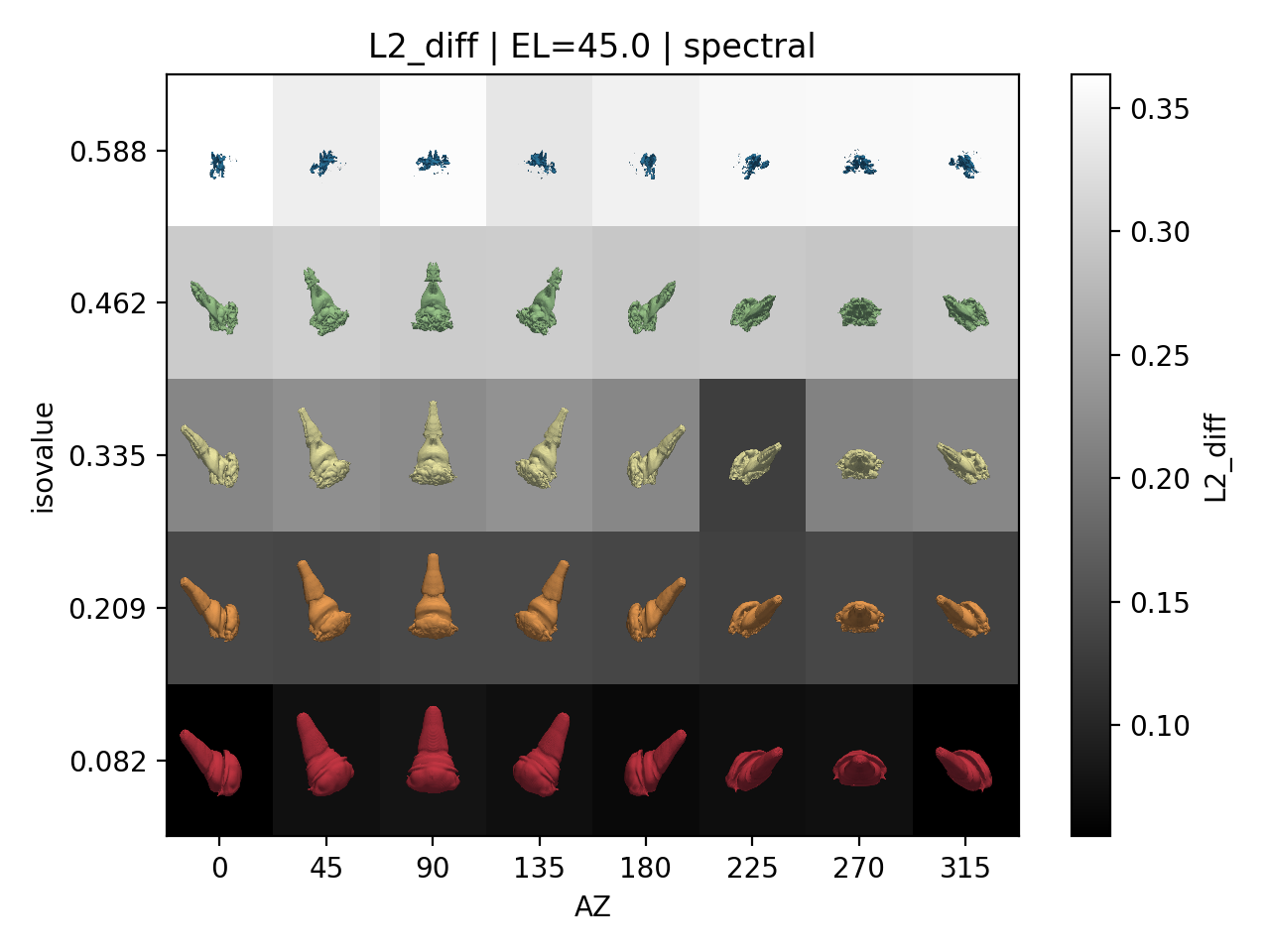}
    }\hfill
\subcaptionbox{}{
        \includegraphics[width=0.47\linewidth]{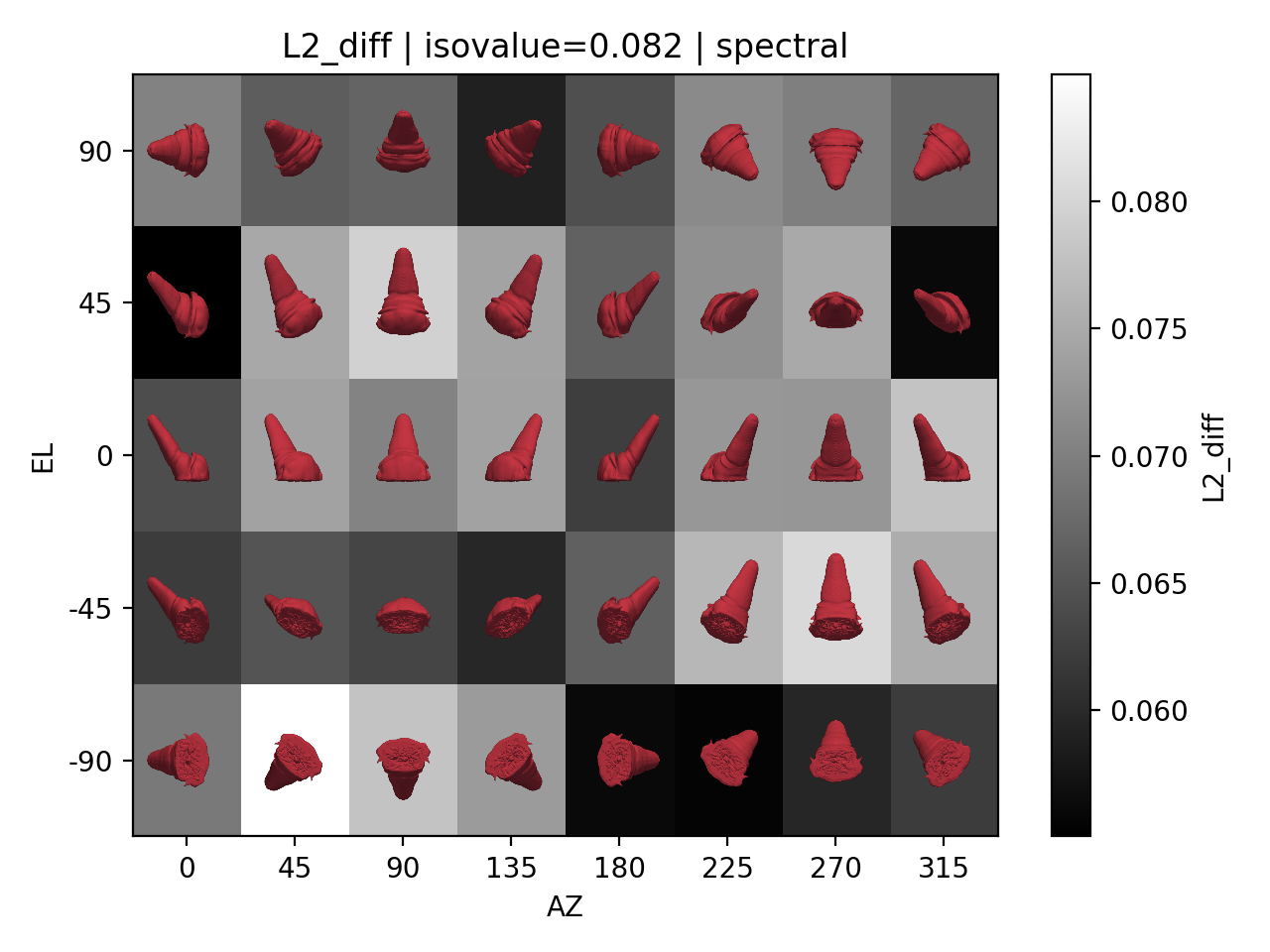}
    }\hfill
     \caption{The reconstruction error depending on the parameters visualized as 2D slices through the global optimum in all pairs of the four parameters isovalue, colormap, azimuth, and elevation. The background color encodes the L2 difference to the original using the greyscale colormap. The foreground shows the central image of the isocontour from the respective angle in the respective colormap.}
    \label{fig:asteroid}
\end{figure}

The data used in this evaluation is the \textit{yA31} All Scalars Asteroid Impact dataset~\cite{patchett2017deep}. The full dataset is an ensemble of simulations of an asteroid impacting the ocean. We use a single timestep of the yA31 simulation output (timestep 52 of 475), which is shortly after the asteroid impacts the ocean, for the visualization. The visualization is a color-coded isocontour of the scalar field \emph{tev}, which represents temperature in electron volts. We generated sets of nine images for the product of five isocontours evenly distributed across the full range $[0.019, 0.652]$ of the variable, i.e., $(0.082, 0.209, 0.335, 0.462, 0.588)$, four colormaps (\texttt{cool-warm}, \texttt{Spectral}, \texttt{viridis}, and \texttt{rainbow}), five elevations ($-90^\circ,-45^\circ,0^\circ,45^\circ,90^\circ$) and eight azimuths ($0^\circ,45^\circ,90^\circ,135^\circ,180^\circ, 225^\circ, 270^\circ, 315^\circ$). This results in a total of 1000 different visualizations.
The best isocontour was detected at $0.082$. 
The best performer of the four colormaps was \texttt{Spectral}. It should be mentioned that this is not a statement about the quality of the whole colormap if we look at a single contour drawn in a single color from that colormap in this case. The best camera angle combination was azimuth $0^\circ$, elevation $45^\circ$. We visualize the reconstruction results in Fig.~\ref{fig:asteroid}. We have a 4D parameter space, which is non-trivial to visualize. We show all 2D slices through the global optimum, comparing each pairwise combination of parameters. 

\begin{figure}[ht]
\subcaptionbox{Isocontour at $0.082$ of the original \emph{tev} field.}{\includegraphics[width=0.49\linewidth]{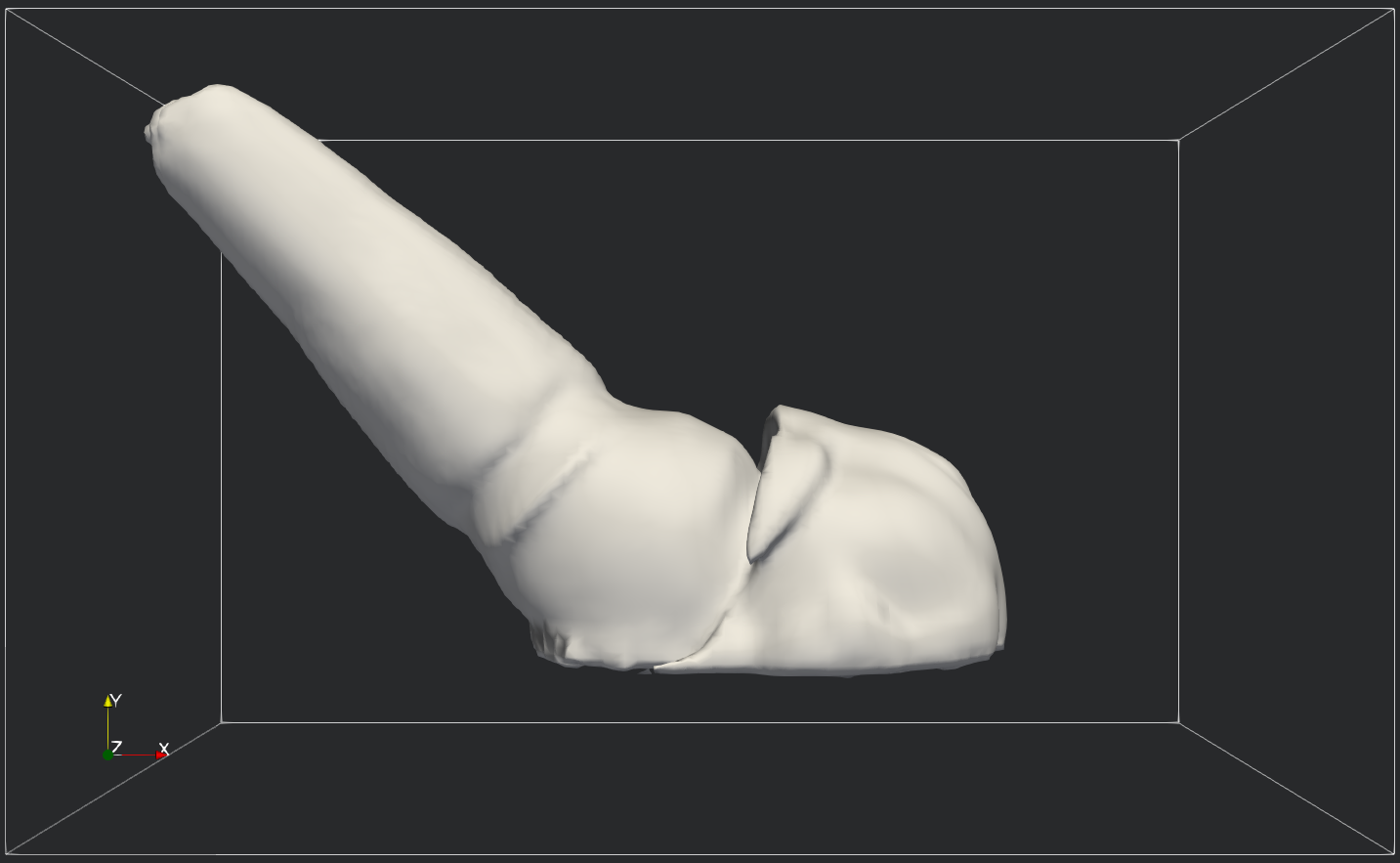}}\hfill
\subcaptionbox{Reconstructed contour.}{\includegraphics[width=0.49\linewidth]{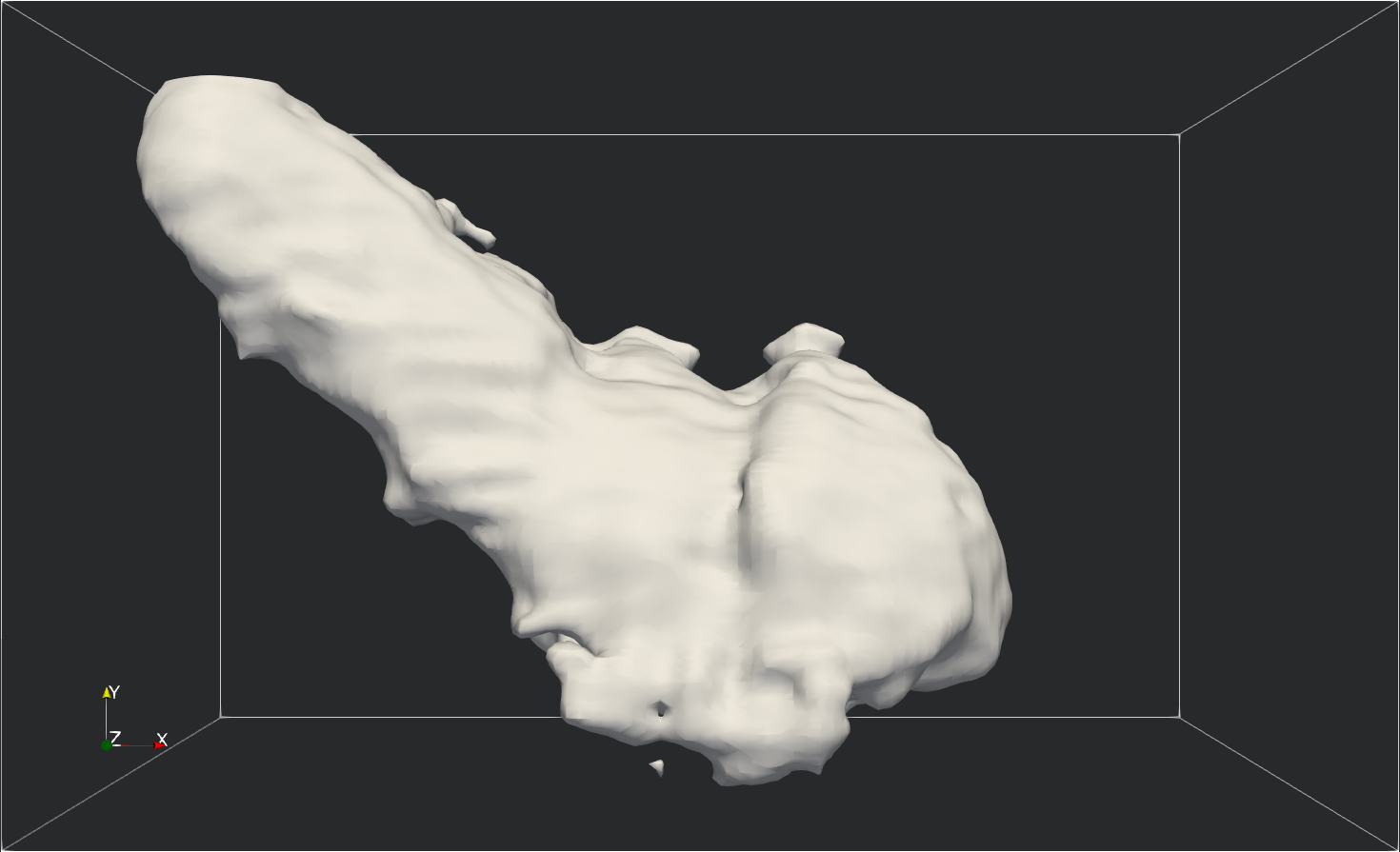}}\hfill
\subcaptionbox{Color-coded slice through the original \emph{tev} field.}{\includegraphics[width=0.49\linewidth]{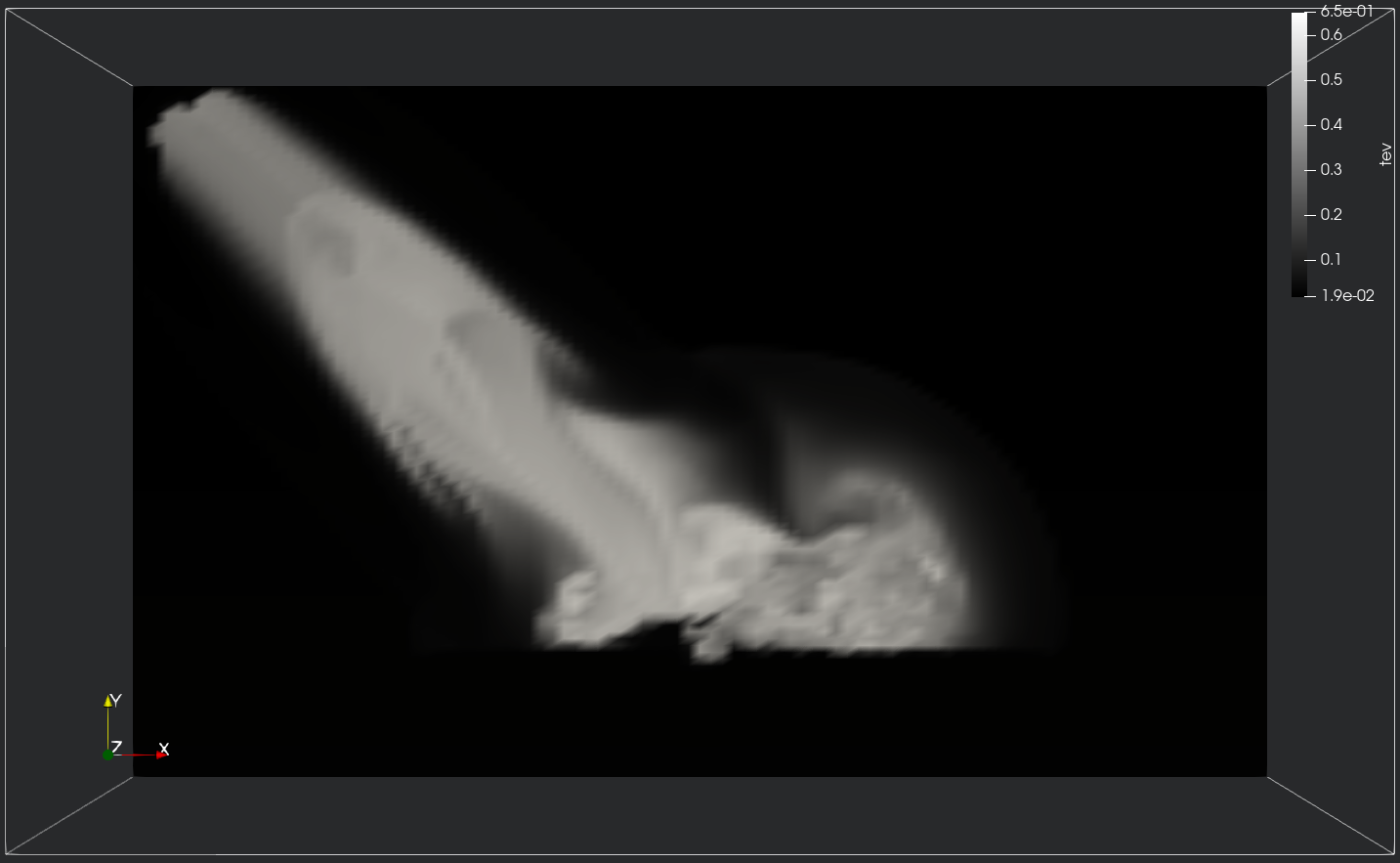}}\hfill
\subcaptionbox{Color-coded slice through the reconstructed field.}{\includegraphics[width=0.49\linewidth]{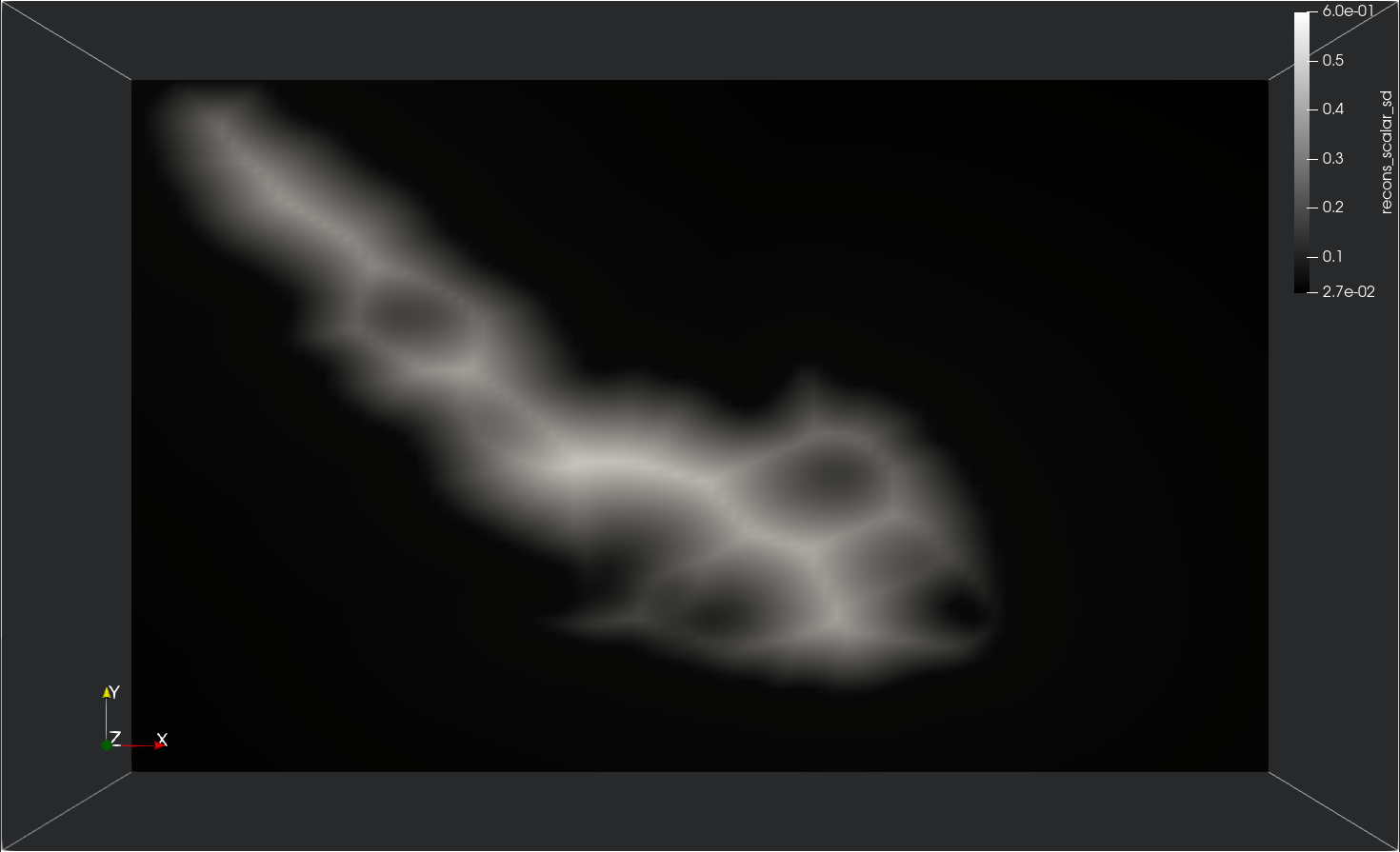}}\hfill
    \caption{Visualization of the original and reconstructed contour and reconstructed volumetric scalar field from the side view.}
    \label{fig:asteroid2}
\end{figure}

We can see a clear preference for the smallest isovalue, which produces the largest contour here. Even though the individual Hausdorff or chamfer distances of other contours might be better, the volumetric reconstruction of the temperature field benefits from having many reference locations. More importantly, as Fig.~\ref{fig:asteroid2} shows, the distribution of the temperature is very non-linear. The background temperature is homogeneous, almost constant, in the air and the water. It jumps abruptly to a high value along the path of the asteroid. Therefore, using a small isovalue that forms a closed hull around the asteroid and its wake produces a very good approximation of the background, which has a large effect on the L2 accuracy. This behavior can be seen as a limitation of the L2 norm: it treats all regions of the domain the same without weighting foreground regions as more important, which a human would~\cite{matzen2017data, mairena2022emphasis}. As mentioned before, the goal of this work is to show that this approach is promising and lay the groundwork for improving it using all the knowledge about visualization. Considering the very non-linear behavior of the temperature in the asteroid dataset and the ill-positioned reconstruction in general, we find that the volumetric reconstruction is actually reasonably good, Fig.~\ref{fig:asteroid2}.

\section{Conclusions}
\subsection{Summary}
We have showcased the power of reconstruction accuracy as a visualization quality metric, which enables self-improving agentic AI workflows to autonomously explore visualization parameter spaces and select the visualization that best represents the underlying data. We have demonstrated its effectiveness through scientific visualization examples, demonstrating its ability to evaluate and optimize visualizations without human intervention as a first, proof-of-concept step.

\subsection{Discussion}
The principle that visualization should faithfully represent the underlying data is understood to be foundational. 
 Ware states that visualization used to refer \emph{``to the act of constructing a visual image in the mind''} and has become \emph{``the creation of graphical representations of data''} by transforming data into a form that facilitates human understanding~\cite{ware2019information}. Similarly, Tufte emphasizes that \emph{``graphics reveal data''}~\cite{tufte1983visual}, underscoring that a visualization’s fundamental goal is to truthfully communicate structure, relationships, and insights inherent in the data. Chen et al.\ extend this perspective, arguing that the primary function of visualization is not just to provide insight but to ensure that insight is grounded in a truthful representation of the data~\cite{chen2014visualization}. 
This notion has been supported by numerous authors~\cite{kindlmann2014algebraic, janicke2011visual, bramon2013information}. 
While the theoretical foundations provided by these authors have long established that visualization should preserve the underlying data, practical implementations have often been constrained by available technology. Most previous approaches have relied on heuristic quality metrics. Though useful in specific contexts, these metrics often serve as indirect proxies for what truly matters: the degree to which the visualization faithfully conveys the data. 

With the advent of NeRF, we now have the computational tools to directly test this principle. NeRF allows us to reconstruct high-fidelity representations of the original data from the visualization itself, effectively mimicking the way the human brain forms a \emph{mental map} from visual input. This represents a paradigm shift: instead of relying on heuristic visual quality metrics, we can now quantitatively assess visualization quality by evaluating how well the data can be recovered from the visualization. 

We have demonstrated that this method provides a more fundamental and reliable measure of visualization effectiveness than existing automated quality metrics. By aligning visualization evaluation more closely with the way humans derive meaning from visual representations, our approach advances visualization research towards more principled, data-grounded assessment methods.

While the state of the art in machine-learned data reconstruction is impressively good, we can expect it to get even better in the future. We therefore believe that this metric is a promising path to extending other AI-based visualization workflows.

\subsection{Limitations and Future Work}
We have already pointed out many limitations of our first order approximation of the final vision in Section~\ref{s:disclaimer}. We hope to incrementally improve all of the int he future. 
Most urgently, we will generalize the reconstruction pipeline to be able to handle other types of data and other visualization metaphors, such as streamlines for vector fields, and glyphs for tensor fields. Currently, we only use machine learning for the reconstruction of the geometry of the visualization metaphors and then use heuristics based on mathematical theory for the reconstruction of the 3D field. We will also explore if incorporating them into the loss function directly in NeRF can force it to directly choose more reasonable values in the non-visible regions. Additionally, we will use reconstruction techniques based on generative AI that are able to mimic the geometric knowledge about the real world to reconstruct 3D scenes from a single image~\cite{bar2024lumiere,gao2024cat3d,  xiang2025structured}. 
Our examples are first-order approximations of the true human mental map. We intend to explore advanced reconstruction algorithms that are trained on perceptual experiments.
The runtimes for the full cross product are currently too long, with NeRF taking about one minute per camera angle. We hope to explore smarter iterative parameter space coverage, maybe making use of the gradients or using faster image-based proxies to decide on camera angles for which a full scene reconstruction is worth computing~\cite{jin2024lvsm}.
Finally, we focused only on data exploration. Extension to task-based visualizations is needed in the future. 

\section*{Acknowledgments}%
	We thank Kei Davis for his feedback on this manuscript. This work is published under LANL LA-UR-25-22935, and was supported by the National Nuclear Security Administration (NNSA) Advanced Simulation and Computing (ASC) Program under grant number DE-SCL0000089. ChatGPT5 was used in the writing of this paper, in particular for language correction.

\FloatBarrier
\bibliographystyle{abbrv}

\bibliography{refs}

\end{document}